\documentclass[fleqn,usenatbib]{mnras}

\usepackage{newtxtext,newtxmath}
\usepackage[T1]{fontenc}

\DeclareRobustCommand{\VAN}[3]{#2}
\let\VANthebibliography\thebibliography
\def\thebibliography{\DeclareRobustCommand{\VAN}[3]{##3}\VANthebibliography}

\usepackage{graphicx}	% Including figure files
\usepackage{amsmath}	% Advanced maths commands
\usepackage{hyperref}
\usepackage{orcidlink}

\title[Dust attenuation in star forming regions]{The IRX--\(\beta\)
Relation in kpc-sized Star Forming Regions in Nearby Galaxies}

\author[L. Duffy et al.]{
Laura Duffy$^{1}$\thanks{E-mail: lrd48@psu.edu}\orcidlink{0000-0003-1752-679X},
Mallory Molina$^{2,3,4}$\orcidlink{0000-0001-8440-3613},
Michael Eracleous$^{1}$\orcidlink{0000-0002-3719-940X},
Robin Ciardullo$^{1}$\orcidlink{0000-0002-1328-0211},
Renbin Yan$^{5}$\orcidlink{0000-0003-1025-1711},
\newauthor
Caryl Gronwall$^{1}$\orcidlink{0000-0001-6842-2371},
Nikhil Ajgaonkar$^{6}$\orcidlink{0000-0003-1469-8246},
M\'ed\'eric Boquien$^{7}$\orcidlink{0000-0003-0946-6176},
Shuang Zhou$^{8}$
and Cheng Li$^{9}$\orcidlink{0000-0002-8711-8970}
\\
$^{1}$Department of Astronomy \& Astrophysics and Institute for Gravitation and the Cosmos, The Pennsylvania State University,\\
\phantom{2}525 Davey Lab, University Park, PA 16803, USA\\
$^{2}$eXtreme Gravity Institute, Department of Physics, Montana State University, Bozeman, MT 59717, USA\\
$^{3}$Department of Physics \& Astronomy, University of Utah, James Fletcher Building, 115 1400 E, Salt Lake City, UT 84112, USA\\
$^{4}$Department of Physics \& Astronomy, Vanderbilt University, Nashville, TN 37235, USA\\
$^{5}$Department of Physics, The Chinese University of Hong Kong, Shatin, N.T., Hong Kong, China\\
$^{6}$Department of Physics and Astronomy, University of Kentucky, 505 Rose St., Lexington, KY 40506-0057, USA\\
$^{7}$Instituto de Alta Investigación, Universidad de Tarapacá, Casilla 7D, Arica, Chile\\
$^{8}$School of Physics \& Astronomy, University of Nottingham, University Park, Nottingham, NG7 2RD, UK\\
$^{9}$Department of Astronomy, Tsinghua University, Beijing 100084, China
}

\date{Accepted XXX. Received YYY; in original form ZZZ}

\pubyear{2015}

\begin{document}
\label{firstpage}
\pagerange{\pageref{firstpage}--\pageref{lastpage}}
\maketitle

\begin{abstract}
The effect of dust attenuation on a galaxy's light depends on a number of physical properties, such as geometry and dust composition, both of which can vary across the faces of galaxies. To investigate this variation, we continue analysis on star-forming regions in 29 galaxies studied previously. We analyse these regions using \textit{Swift}/UVOT and WISE images, as well as SDSS/MaNGA emission line maps to constrain the relationship between the infrared excess (IRX) and the UV spectral index, $\beta$ for each star forming region. This relationship can be used to constrain which dust attenuation law is appropriate for the region. We find that the value of D$_n(4000)$ for a region is correlated with both IRX and $\beta$, and that the gas-phase metallicity is strongly correlated with the IRX. This correlation between metallicity and IRX suggests that regardless of aperture, metal rich regions have steeper attenuation curves. We also find that integrated galactic light follows nearly the same IRX-$\beta$ relationship as that found for kiloparsec-sized star forming regions. This similarity may suggest that the attenuation law followed by the galaxy is essentially the same as that followed by the regions, although the relatively large size of our star forming regions complicates this interpretation because optical opacity and attenuation curves have been observed to vary within individual galaxies.
\end{abstract}

\begin{keywords}
dust, extinction -- galaxies: star formation -- galaxies: ISM
\end{keywords}

%%%%%%%%%%%%%%%%%%%%%%%%%%%%%%%%%%%%%%%%%%%%%%%%%%

%%%%%%%%%%%%%%%%% BODY OF PAPER %%%%%%%%%%%%%%%%%%

\section{Introduction}
The accurate determination of galactic star formation rates (SFRs) is crucial in the study of galaxy evolution. Rest frame near ultraviolet (UV) emission is one of the most common tracers of recent star formation \citep{kennicuttevans}. This is especially true at high redshift where rest-frame UV is more easily accessible than rest-frame infrared (IR). Observations of star formation are often complicated by the production of dust, which hides newly formed stars. In fact, it has been shown that the fraction of star formation hidden by dust is dominant at redshifts $z\sim1-3$ \citep{lefloch, magnelli, reddy}. Furthermore, UV light is preferentially extinguished by dust. Our ability to correct for the effect of dust depends heavily upon our knowledge of the appropriate extinction and attenuation laws. To learn how to correct for these effects, we often rely on nearby systems.

Attenuation laws attempt to quantify the effects that dust, of varying composition and geometrical configuration, has on the propagation of light \citep[see][for a review]{salimnar}. They differ from extinction laws in that they account for the geometrical complexities that exist in physical systems. They are generally parameterized using the same two basic properties \citep{noll}. The first is the general shape of the curve as a function of wavelength. The second is the strength of the excess attenuation at 2175~\AA, referred to as the 2175~\AA\ bump. Both correlate with the intrinsic physical properties of the object being observed, such as the SFR and total stellar mass \citep{calzetti00, wild11, salim18}, as well as dust composition, geometry and the V-band attenuation \citep[$A_V$; e.g.][]{boquien22}.

The physical properties that affect the shape of the attenuation law can vary both from galaxy to galaxy as well as across a single galaxy \citep{charlotfall, calzetti00, wild11}. The fact that dust extinction laws within a galaxy vary greatly from sight-line to sight-line \citep[e.g.,][]{cardelli, fitzpatrick, gordon} means that the attenuation curve will also vary, further complicating the measurements of small-scale star formation \citep{hoversten, hagen, 2020MNRAS.494.4751M, zhou22}.

Although it is common to apply attenuation corrections when fitting spectral energy distributions (SEDs), such corrections become difficult when photometry is only available in a few bands. When these situations arise, corrections can be obtained by using a combination of attenuation-sensitive star formation tracers and the amount of emission in the IR. One such scheme involves the relationship between the UV spectral index, $\beta$, and the infrared excess \citep[IRX;][]{Calzetti94, meurer99, calzetti00}. This scheme requires assuming that all the light emitted by young stars in the UV that is attenuated by dust is then re-emitted in the IR, and that there is no sight-line dependence on the UV attenuation \citep{dacunha}. This relationship becomes especially useful at high redshift where rest-frame IR light becomes difficult to observe, and correction based solely on rest-frame UV light might be necessary.

The IRX is defined as $\mathrm{IRX}\equiv \log(L_\mathrm{TIR}/L_\mathrm{FUV})$, where $L_\mathrm{TIR}$ refers to the total IR luminosity of the source, often measured between 3 and 1100 $\mu$m, and $L_\mathrm{FUV}$ is the monochromatic luminosity at some far-ultraviolet (FUV) wavelength, often that of the GALEX FUV filter \citep{meurer99}. The IRX thus provides a measure of the amount of light initially emitted in the UV that is absorbed and re-emitted by dust in the IR, and is, in effect, a good tracer of the attenuation.

The UV spectral index, $\beta$, quantifies the shape of the SED of a galaxy in the UV band by assuming that the stellar continuum is well approximated by $F_\lambda \propto \lambda^\beta$ \citep{Calzetti94}.The relationship means that negative values of $\beta$ are characteristic of very young populations with little dust, and positive values of $\beta$ correspond to older or dustier populations \citep[see][]{calzetti_review}.

The relationship between IRX and $\beta$ has been studied for both starburst \citep[e.g.][]{meurer99, kong04, takeuchi12, nagaraj} and local star-forming galaxies \citep[e.g.][]{dale09, takeuchi12, salimboquien19}, as well as on sub-galactic scales across star-forming galaxies \citep[e.g.][]{boquien09, boquien}. \cite{meurer99} first defined an IRX-$\beta$ relationship for starburst galaxies. Later measurements of the relationship for more normal star-forming systems, however, found deviations from the \cite{meurer99} law. For example, \cite{boquien} inspected seven, face-on, star-forming galaxies on a pixel-by-pixel basis and found that deviations from the \cite{meurer99} relationship could be caused either by variations in the intrinsic UV colour of the underlying stellar populations or by varying attenuation law slopes. Alternatively, \cite{takeuchi12}, exploring the same deviation, found that the sample of starburst galaxies studied by \cite{meurer99} did, in fact, follow the same relationship found for normal star-forming galaxies, but that aperture effects played a large role in shifting the apparent IRX-$\beta$ relationship.

While previous work has investigated relationships between the IRX and $\beta$ on subgalactic scales \citep{munoz09, boquien, ye}, those studies were limited by the number of galaxies with resolved regions. We attempt to understand and explain the drivers of attenuation in those regions, and the relationship between attenuation in resolved regions and attenuation in integrated galactic light in a larger sample of 29 galaxies. Our work is a continuation of that started by \cite{2020MNRAS.494.4751M}, which focused on the source of variation in the attenuation laws by using the optical and near UV tracers $\tau^l_{\rm B}$, or Balmer optical depth, and $\beta$.

We combined data from the \textit{Swift}/UV Optical Telescope (UVOT), the Mapping Nearby Galaxies at Apache Point Observatory (MaNGA) integrated field unit (IFU) survey, and the Wide-field Infrared Survey Explorer (WISE) for 29 galaxies examined by \cite{2020MNRAS.494.4751M} to study dust attenuation via the IRX-$\beta$ relationship in resolved, kpc-sized star forming regions. These galaxies were drawn from a larger catalogue, created to study star formation and its quenching in the local universe \citep{swim}. We also compared the resulting relationship with that for the integrated galactic light, and examined the role of a number of physical parameters in driving the scatter of the relationship.

In Section \ref{sec:sample}, we introduce the sample. Section \ref{sec:process} describes the steps that were taken to process the data. We introduce the methodology behind defining and analysing the star forming regions in Section \ref{sec:starform}. In Section \ref{sec:analysis}, we investigate the effect that observed galaxy properties have on the relationship between IRX an $\beta$. We discuss and conclude upon our findings in Section \ref{sec:conc}. We assume a $\Lambda$CDM cosmology, with $\Omega_\mathrm{m}=0.3$, $\Omega_\Lambda = 0.7$ and $H_0=70\;\mathrm{km}\;\mathrm{s^{-1}}\;\mathrm{Mpc}^{-1}$.

\section{Samples \& Dataset}\label{sec:sample}

\subsection{Target Selection}
\label{sec:targ} % used for referring to this section from elsewhere
This work focuses on a subset of galaxies drawn from the larger \textit{Swift}/MaNGA Value Added Catalogue \citep[SWIM VAC; ][]{swim}, created to study star formation and its quenching in nearby galaxies. The SwiM VAC consists of \textit{Swift}/UVOT images, a wide variety of MaNGA emission line maps (including indices measured from MaNGA spectra), and Sloan Digital Sky Survey (SDSS) images. \cite{2020MNRAS.494.4751M} used the catalogue to investigate the variation in attenuation laws for resolved, kpc-sized star forming regions. Those authors selected 29 of the original 150 SwiM VAC galaxies with appropriate star formation rates and properties and used reddening-insensitive emission line ratios to identify galaxies where the majority of pixels were star-forming according to the three standard emission line diagnostic diagrams, commonly referred to as the BPT diagrams \citep{bpt, kauffmann03, kewley06}. Here, we further explore dust attenuation in those 29 galaxies. The basic properties of the sample galaxies can be found in Table \ref{table:properties}. The galaxies have a median redshift of $z_m = 0.03$, and have masses that span the range $8.8\leq \log(M/M_\odot)\leq 10.7$, similar to the sample used in \cite{B16}.

\begin{table*}
%\centering
\caption{Basic properties of the sample of star forming galaxies}\label{table:properties}
\setlength{\tabcolsep}{10 pt}
\begin{tabular}{cllcccccc}
\hline\hline
 Galaxy   &  Galaxy   &  SDSS  & & & & $R_{pet}$ & $\log(M_*)$ & $\log(\mathrm{SFR})$ \\
 I.D. & Name & Class & $z$ & $E(B-V)_G$ & $b/a$ & (arcsec) & ($M_\odot$) & ($M_\odot\:\mathrm{yr}^{-1}$)\\
 (1) & (2) & (3) & (4) & (5) & (6) & (7) & (8) & (9) \\
\hline
1 & UGC 11696 & SF & 0.017 & 0.101 & 0.52 & 19 & 8.76 & $-1.23$\\
2 & SDSS J170653.67+321010.1 & SF & 0.036 & 0.039 & 0.82 & 17 & 9.15 & $-0.68$\\
3 & WISE J162856.63+393634.1 & SF & 0.035 & 0.010 & 0.48 & 16 & 9.19 & $-0.62$\\
4 & 2MASS J04072365-0641117 & SF & 0.038 & 0.098 & 0.60 & 16 & 9.25 & $-0.71$\\
5 & 2MASX J11044509+4509238 & SF & 0.022 & 0.008 & 0.88 & 16 & 9.27 & $-1.52$\\
6 & 2MASS J14135887+4353350 & SB & 0.040 &  0.013 & 0.71 & 12 & 9.31 & $-0.61$\\
7 & KUG 1016+468 & SF & 0.024 &  0.013 & 0.45 & 19 & 9.32 & $-0.87$\\
8 & SDSS J030659.79-004841.5 & SF & 0.038 & 0.070 & 0.78 & 16 & 9.40 & $-0.61$\\
9 & 2MASS J14132780+4354501 & SF & 0.040 & 0.013 & 0.40 & 16 & 9.43 & $-0.56$\\
10 & SDSS J024112.93-005236.9 & SF & 0.038 & 0.034 & 0.56 & 19 & 9.44 & $-1.07$\\
11 & 2MASS J15004786+4836270 & SF & 0.037 & 0.020 & 0.54 & 14 & 9.50 & $-0.54$\\
12 & 2MASXi J0740537+400411 & N/A & 0.042 & 0.052 & 0.59 & 18 & 9.50 & $0.47$\\
13 & KUG 0757+468 & SF & 0.019 & 0.065 & 0.53 & 19 & 9.52 & $-0.03$\\
14 & 2MASX J14403849+5328414 & SB & 0.038 & 0.011 & 0.80 & 15 & 9.55 & $-0.12$\\
15 & KUG 0254-004 & SF & 0.029 & 0.066 & 0.80 & 17 & 9.65 & 0.00\\
16 & 2MASX J07522873+4950192 & SF & 0.022 & 0.059 & 0.50 & 20 & 9.67 & $-0.86$\\
17 & KUG 0751+485 & SB & 0.022 & 0.035 & 0.71 & 15 & 9.71 & $-$0.46\\
18 & 2MFGC 8582 & SF & 0.025 & 0.014 & 0.33 & 18 & 9.73 & $-0.23$\\
19 & 2MASX J07595878+3153360 & SB & 0.045 & 0.048 & 0.57 & 16 & 9.78 & 0.08\\
20 & KUG 1121+239 & SF & 0.028 & 0.019 & 0.80 & 21 & 9.79 & $-0.45$\\
21 & WISE J163226.31+393104.0 & SF & 0.029 & 0.009 & 0.87 & 17 & 10.0 & $-0.53$\\
22 & 2MASX J09115605+2753575 & SB & 0.047 & 0.028 & 0.48 & 15 & 10.1 & 0.17\\
23 & KUG 1343+270 & SF & 0.030 & 0.017 & 0.54 & 22 & 10.1 & $-0.22$\\
24 & 2MASX J11532094+5220438 & SF & 0.049 & 0.029 & 0.87 & 16 & 10.1 & 0.00\\
25 & KUG 1626+402 & SF & 0.026 & 0.009 & 0.70 & 20 & 10.2 & 0.15\\
26 & NGC 3191 & SF & 0.031 & 0.011 & 0.86 & 22 & 10.3 & 0.73\\
27 & LCSB S1611P & Galaxy & 0.056 & 0.010 & 0.40 & 16 & 10.4 & $-0.13$\\
28 & 2MASX J07333599+4556364 & SF & 0.077 & 0.091 & 0.89 & 17 & 10.5 & 0.94\\
29 & 2MASX J03065213-0053469 & SF & 0.084 & 0.070 & 0.88 & 17 & 10.7 & 0.56\\
\hline
\end{tabular}
Column 1: Galaxy ID number in this sample. Column 2: Galaxy name. Column 3: Galaxy class from \cite{bolton}. SF stands for star forming, SB stands for starburst. Column 4: Redshift from the NASA Sloan Atlas \citep{sdssiv}. Column 5: Foreground extinction measurements for each galaxy from \cite{1998ApJ...500..525S}. Column 6: Ratio of minor to major axis from the NASA Sloan Atlas \citep{sdssiv}. Column 7: Petrosian radii are calculated in the W3 band following the procedure outlined in Section \ref{sec:skysub}. Column 8: Stellar masses are taken from SED fits from the MaNGA Data Reduction Pipeline \citep{law16}. Column 9: SFRs using the H$\alpha$ luminosity are from the MaNGA Data Analysis Pipeline \citep{westfall19}.

\end{table*}

Galaxy I.D. 26 (NGC 3191) from the original sample of \cite{2020MNRAS.494.4751M} is excluded from most of our analyses, as the \textit{Swift}/UVOT imaging of the galaxy coincided with the appearance of a Type I superluminous supernova in its central parts \citep[see][]{ngc3191}. Since observations in the other bands were not contemporaneous, comparisons between the MaNGA emission line maps, \textit{Swift} UV observations, and WISE images cannot be used to produce a reliable value of $\beta$.

\subsection{SDSS-IV/MaNGA and Swift/UVOT}
\cite{swim} constructed the SwiM VAC by matching galaxies in the first data release of MaNGA fully reduced spectra \citep[DRP;][]{law16, MaNGA_bundy, MaNGA_r, aguado19} to \textit{Swift}/UVOT observations. MaNGA is one of the three major surveys in SDSS-IV \citep{sdssiv}, and utilizes hexagonal bundles of $2''$ fibers that are mounted directly on the Sloan 2.5~m telescope \citep{gunn06} and are fed into the Baryon Oscillation Spectroscopic Survey (BOSS) spectrographs \citep{smee13, drory15}. The MaNGA survey aimed to target $\sim$ $10,000$ galaxies with uniform spatial coverage and a uniform distribution in stellar mass for M$_*$ $>$ $10^9$ $\mathrm{M}_\odot$, while also maximising spatial resolution and signal-to-noise ratio (S/N) for each galaxy \citep{manga_sample}. The effective point-spread function (PSF) of the resulting maps has a full width at half maximum (FWHM) of $2\farcs5$ \citep{law15, law16}. MaNGA spectra have a resolving power $R\sim 2000$, and cover a wavelength range $3622-10354$ \AA, which provides many nebular diagnostic lines \citep{westfall19, belfiore19}. MaNGA spectra also have very small errors in their flux calibration \citep{yan16a}.

The \textit{Swift}/UVOT is a 30-cm telescope with an effective plate scale of $1''$ pixel$^{-1}$ \citep{swift_ref}. Here we use the uvw2 and uvw1 images, which have PSF FWHM values of $2\farcs92$ and $2\farcs37$, respectively. Notably, these FWHM values are similar to that of MaNGA\null. The basic properties of the two NUV filters are listed in Table \ref{table:filters}. Each of the star forming galaxies in our sample has a minimum S/N of 15 in both near-UV (NUV) filters for the integrated galaxy light, and our sample galaxies have a median exposure times of 2296 and 2375 seconds in the uvw1 and uvw2 filters, respectively \citep{swim}.

\begin{table}
%\centering
\caption{Basic properties of the \textit{Swift} UVOT uvw1 and uvw2 filters, and the WISE W3 filter.}\label{table:filters}
\setlength{\tabcolsep}{17 pt}
\begin{tabular}{lccc}
\hline\hline
 & Central & PSF & Pixel \\
Filter & Wavelength & FWHM & Size \\
(1) & (2) & (3) & (4) \\
\hline
uvw2 & 1928 \AA & 2\farcs92 & 1\arcsec \\
uvw1 & 2600 \AA & 2\farcs37 & 1\arcsec \\
W3 & 12 $\mu$m & 6\farcs5 & 2\farcs75\\
\hline
\end{tabular}
\\ Column 1: The filter. Column 2: The central wavelength of the filter. Column 3: The full width at half maximum of the point spread function for each filter. Column 4: The pixel size for each filter.
\end{table}

\subsection{WISE}
We supplement the above data set with infrared images for the 29 galaxies from WISE. The WISE telescope, which has a 40-cm primary mirror, mapped the sky in four infrared bands, with filters centred at 3.4, 4.6, 12, and 22 $\mu$m \citep{wise}. The W1, W2, and W3 bands each have a PSF FWHM of about $6''$ whereas the W4 band PSF FWHM is about $12''$.  The W1, W2, and W3 images also have angular pixel sizes of $2\farcs75$ while the W4 images have a pixel size of $5\farcs5$. 

While the 22 $\mu$m W4 band is better suited in wavelength for probing the total infrared luminosity of galaxies, it cannot resolve the kpc-sized star forming regions of interest to this study. For this reason, we do not use W4 band photometry, and instead rely on W3 measurements.

\section{Data Processing}\label{sec:process}

\subsection{Swift/UVOT Data Processing}
We used \textit{Swift}/UVOT count and exposure maps which were pre-processed by \cite{2020MNRAS.494.4751M} with the \textit{Swift}/UVOT Pipeline\footnote{\texttt{github.com/malmolina/Swift-UVOT-Pipeline}} (see section 3.2 of that paper). These count and exposure rates were further processed and presented in \cite{swim}. We created images that were then corrected for the foreground extinction of each galaxy using the average total-to-selective extinction values for uvw1 and uvw2 found by \cite{2020MNRAS.494.4751M}. E(B-V) values were taken from \cite{1998ApJ...500..525S}. We then applied the Fitzpatrick extinction law \citep{fitzpatrick}, chosen by \cite{2020MNRAS.494.4751M} in order to facilitate comparison of their results with \cite{B16}. E(B-V) values for each galaxy can be found in Table \ref{table:properties}. 

\subsection{Sky Subtraction}\label{sec:skysub}
We followed the same procedure for subtracting the background from UVOT images as \cite{2020MNRAS.494.4751M}. To summarize, we first calculated the Petrosian radius of each galaxy in the uvw2 filter. We then constructed a 1000~arcsec$^2$ annulus whose inner radius was twice the Petrosian radius. Any bright objects within the annulus were masked to avoid contamination. Finally, we calculated the background sky counts per pixel in the annulus, using the biweight estimator \citep[see][]{biweight}, and subtracted them from the images. In all cases, the background counts were minimal.

For the WISE W3 images, we followed a slightly modified procedure, as it was impossible to calculate a Petrosian radius without first making a preliminary estimate of the background. Therefore we first selected three small, blank-sky regions far away from bright sources in each image, estimated the counts in each region, averaged the three resulting values, and subtracted this initial background estimate from each pixel. We then calculated the Petrosian radius for each program galaxy in the W3 filter, and followed the procedure laid out in \cite{2020MNRAS.494.4751M} to subtract the background more carefully.  Unlike the UVOT images, in many cases, the background in W3 was substantial.

\subsection{Resolution matching of MaNGA maps, Swift images, and WISE images}
While NUV images from UVOT and emission line maps from MaNGA have similar spatial resolution, the resolution of the WISE W3 images is considerably lower (see Table \ref{table:filters}). Therefore, in order to directly compare the three data sets, the MaNGA emission line maps and the UVOT NUV images were converted to the WISE W3 band angular resolution and pixel size. To do this, we first convolved each of the MaNGA emission line maps and the uvw1 and uvw2 images with an appropriate kernel to match the WISE W3 band PSF.

Following \cite{swim}, we define the convolution kernel as a 2D Gaussian with a standard deviaiton
\begin{equation}
    \sigma = \sqrt{\frac{\mathrm{FWHM}_{\mathrm{W}3}^2 - \mathrm{FWHM}_\mathrm{x}^2}{8\,\mathrm{ln}2}}
\end{equation}
where FWHM$_\mathrm{x}$ represents the FWHM of the PSF of the corresponding filter.

The MaNGA emission line maps from the SwiM VAC and the \textit{Swift} images have a pixel scale of $1''$ whereas WISE W3 band images have a pixel scale of $2\farcs 75$. Therefore, following convolution, we reprojected the MaNGA maps and \textit{Swift} images to the WISE W3 sampling. For this step, we used the \texttt{reproject.reproject\_exact} function from \texttt{astropy} \citep{astropy}, which employs a flux-conserving spherical polygon intersection algorithm. Throughout the process, we masked bad pixels.

\begin{figure*}
\centering
\includegraphics[width=0.45\textwidth]{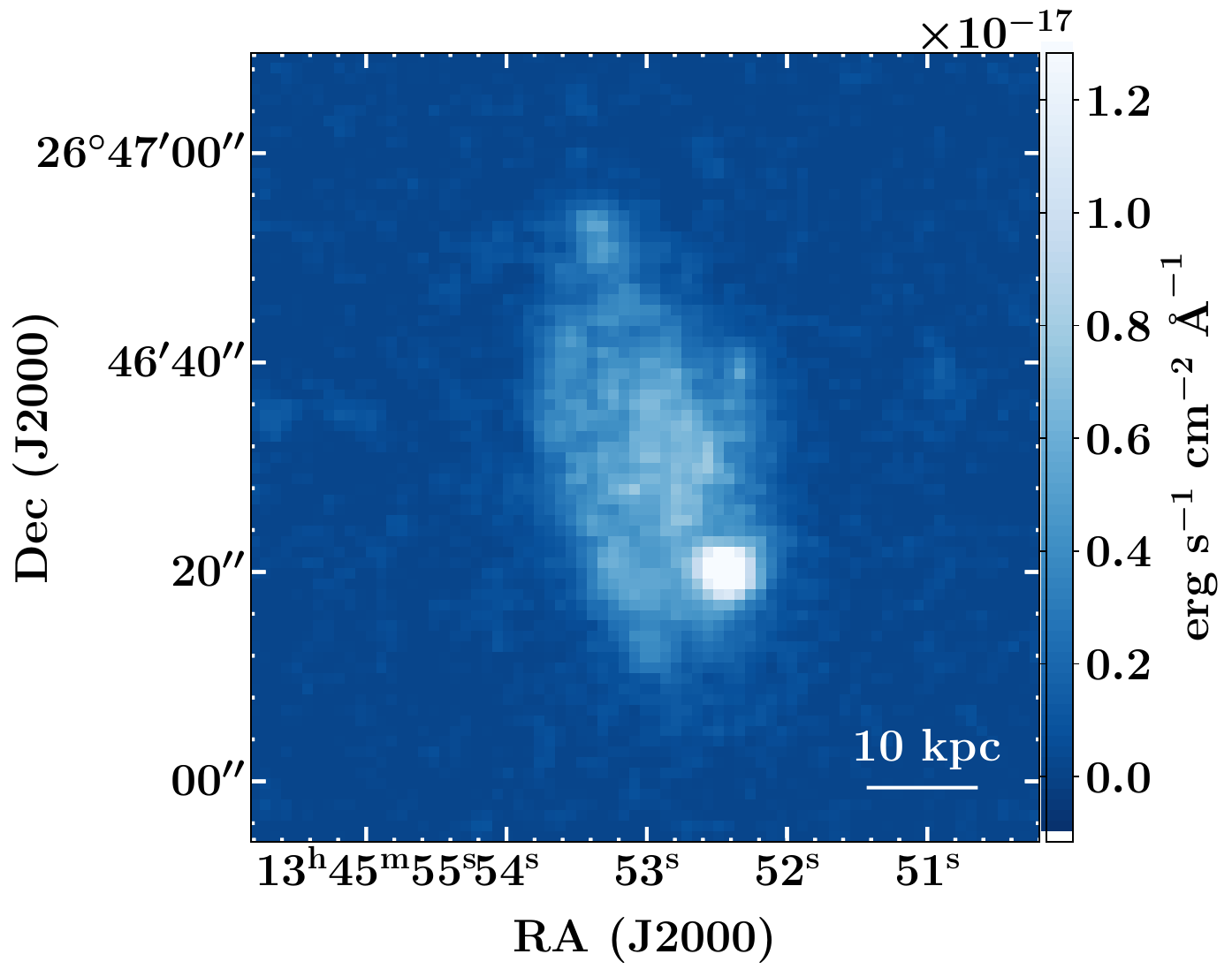}
\hskip 1cm
\includegraphics[width=0.45\textwidth]{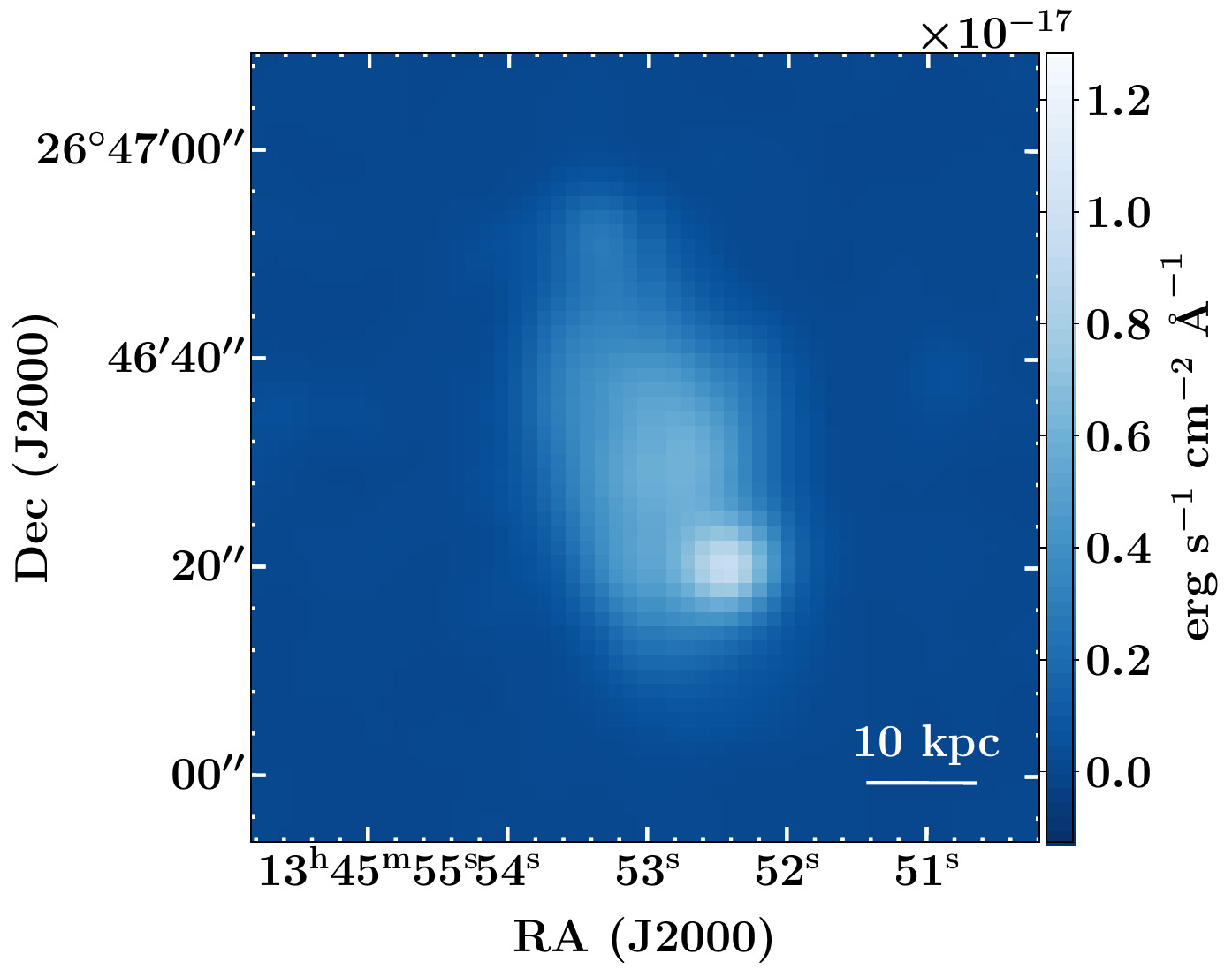}
\caption{The uvw2 filter image of galaxy I.D. 23 (KUG 1343+270) shown before (left) and after (right)  On the left is the galaxy at the original spatial sampling and resolution and on the right is the galaxy at the W3 sampling and resolution. While general morphology still resolvable at both resolutions, the level of detail is noticeably decreased.}
\label{fig:beforeafter}
\end{figure*}

The data processing steps described above caused a substantial change in the resolution of the MaNGA maps and \textit{Swift} UV images. We illustrate this in Figure \ref{fig:beforeafter}, where we show the uwv2 image of galaxy I.D. 23 (KUG 1343+270) before and after convolution. Notably, a great deal of detail is lost after the convolution. Any aperture that is placed around a source in the convolved and re-projected image will have less light in it from the source than the same aperture placed on the original frame. Because all images are at the same resolution after this processing, aperture corrections are not necessary here.

\section{Kpc-Sized Star Forming Regions}\label{sec:starform}

\subsection{Identification of Star Forming Regions}\label{sec:starformid}
\cite{2020MNRAS.494.4751M} detailed the method they used to define the kpc-sized star forming regions in the galaxies that are the subject of our study. However, because our images are degraded to the WISE resolution, we redefined the regions at the new resolution. Following \cite{2020MNRAS.494.4751M}, we used dereddened, convolved, and reprojected H$\alpha$ surface brightness maps in conjunction with emission line diagnostic diagrams and H$\alpha$ equivalent widths to define the new resolved star forming regions.

We adopted a minimum H$\alpha$ surface brightness, $\log[\Sigma_{\mathrm{H}\alpha}/(\mathrm{erg\;s}^{-1}\;\mathrm{kpc}^{-2})] = 39$, to locate areas dominated by light from H II regions. This limit, recommended by \cite{zhang}, helps to identify pixels where H$\alpha$ emission is primarily due to recent star formation, as opposed to light from diffuse ionized gas (DIG\null). We also required that the H$\alpha$ equivalent width, EW(H$\alpha$), be larger than $15$ \AA. EW(H$\alpha$) is a mass-specific proxy for SFR, and is used here to mitigate the effect that diffuse galaxy light has on the identification of star forming regions. These requirements mean that in all regions there should be a fairly high level of recent star formation.

We then followed the procedure of \cite{2020MNRAS.494.4751M}, which we summarize below, to delineate star-forming regions:

\begin{enumerate}
    \item[(i)] We identified local peaks in the surface brightness with \(\mathrm{log}[\Sigma_{\mathrm{H}\alpha}/(\mathrm{erg\:s^{-1}\:kpc^{-2})}]\geq 39\).
    
    \item[(ii)] We defined the largest isophote that corresponds to \(\mathrm{log}[\Sigma_{\mathrm{H}\alpha}/(\mathrm{erg\:s^{-1}\:kpc^{-2})}]\geq 39\). If the contour has a diameter of less than \(12''\), which we set as the minimum allowed aperture size based on the resolution of the images, adopt a circular aperture \(12''\) diameter.
    
    \item[(iii)] We ensured that nebular emission line ratios from the region of interest fall outside of the AGN regions on each of the emission line diagnostic diagrams.
    
    \item[(iv)] We repeated previous steps, shifting the contour levels in increments of \(0.2\) in \(\mathrm{log}[\Sigma_{\mathrm{H}\alpha}]\) and stopping at \(\mathrm{log}[\Sigma_{\mathrm{H}\alpha}/(\mathrm{erg\:s^{-1}\:kpc^{-2})}]= 38\), while ensuring that no region is identified more than once.
    
    \item[(v)] We calculated the diameter of the star forming regions to ensure that they are still kpc-sized.
\end{enumerate}

\begin{figure*}
\centering
\includegraphics[width=0.45\textwidth]{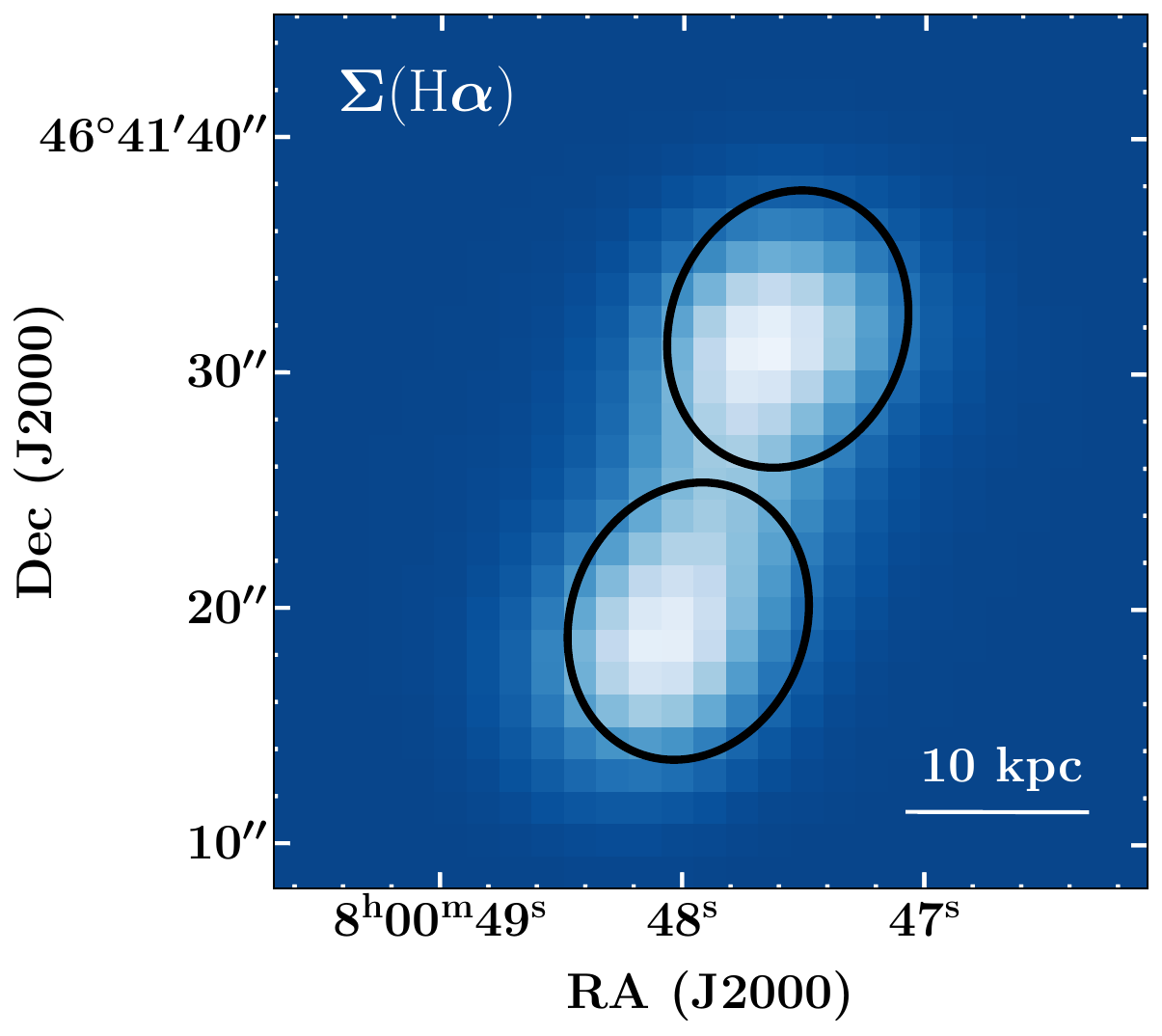}
\hskip 1cm
\includegraphics[width=0.45\textwidth]{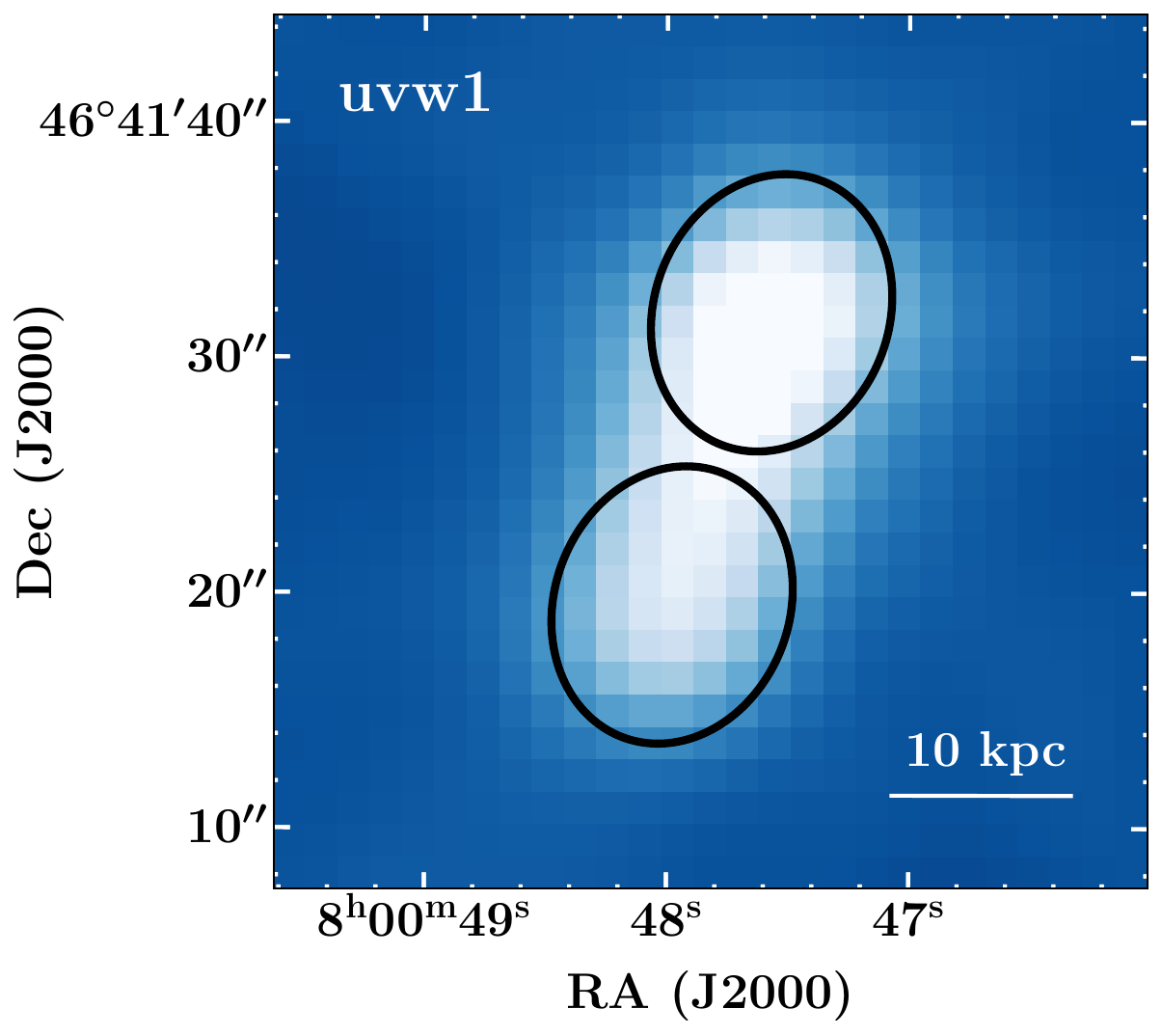}
\vskip 1cm
\includegraphics[width=0.45\textwidth]{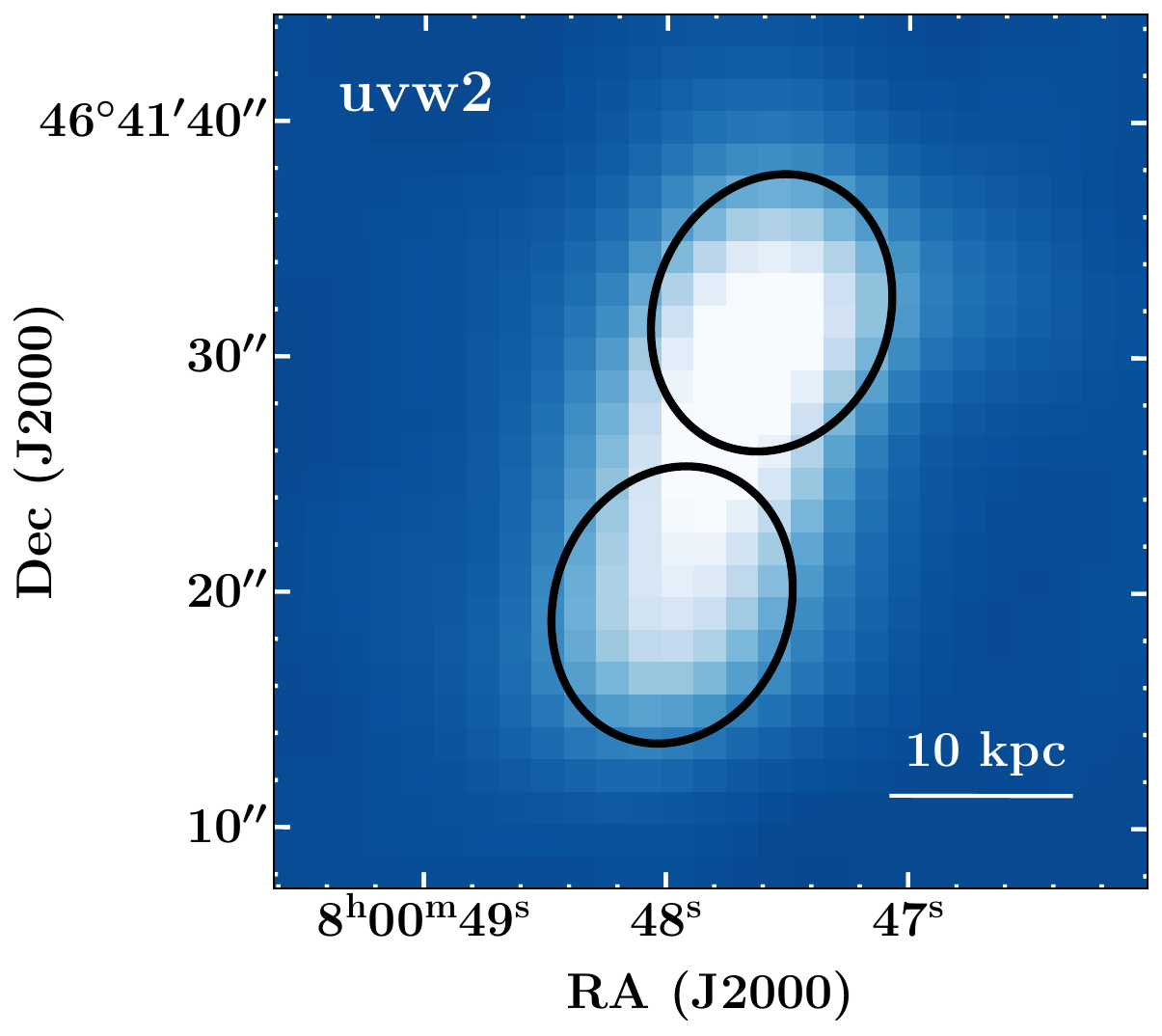}
\hskip 1cm
\includegraphics[width=0.45\textwidth]{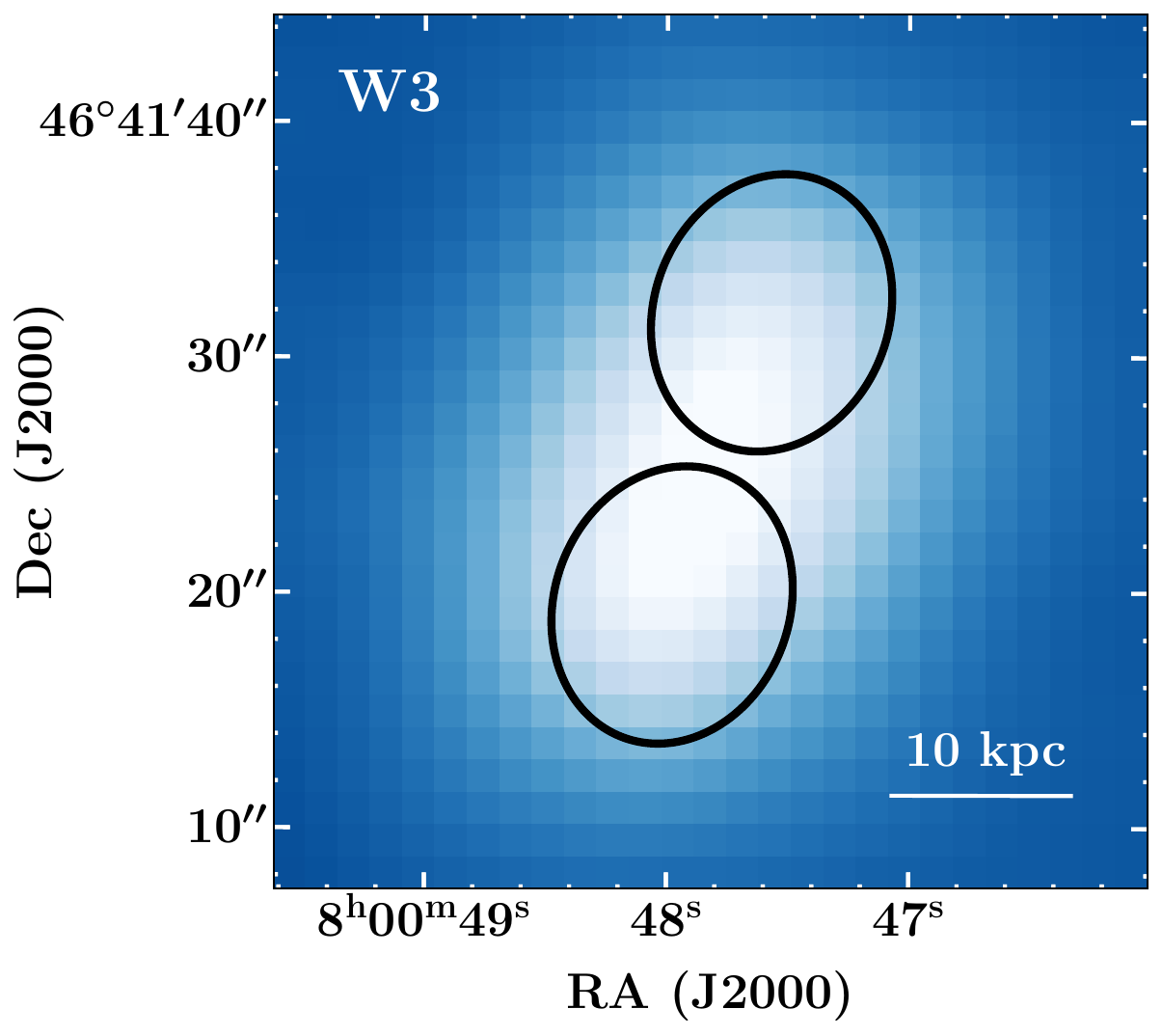}
\caption{H$\alpha$ surface brightness map and uvw1, uvw2 and W3 images of Galaxy I.D. 13 (KUG 0757+468). Each map or image has been convolved and reprojected to the W3 resolution. Outlined in black are the two identified star forming regions in the galaxy.}
\label{fig:allbands}
\end{figure*}

To illustrate the results of this method, in Figure \ref{fig:allbands}, we show images of galaxy I.D. 13 (KUG 0757+468) in the four most relevant bands for our analysis. We outline the two identified star forming regions in black. This figure includes an H$\alpha$ emission line flux map, the UVOT uvw2 and uvw1 images, and the WISE W3 image.

\begin{figure*}
    \centering
    \includegraphics[width=0.6\linewidth]{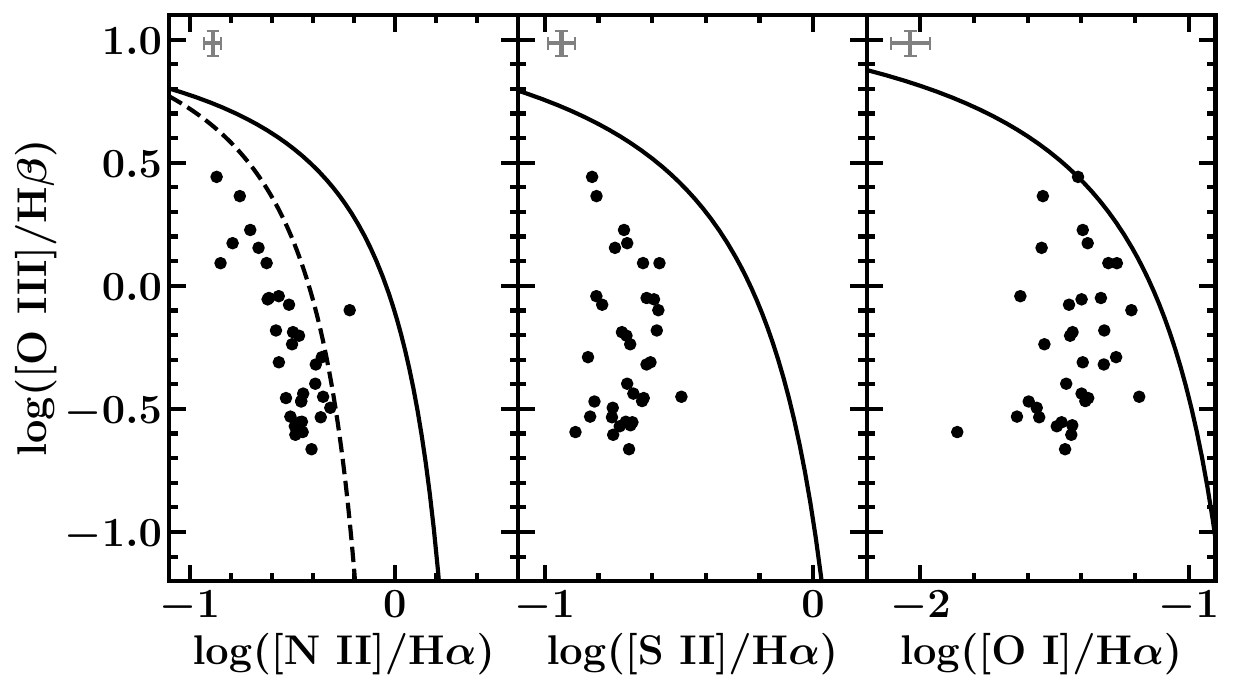}
    \caption{Emission line diagnostic diagrams for individual star forming regions. The solid black line in each panel shows the extreme starburst limit from \citet{kewley06}. The black dashed line represents the limit between the star forming region and the composite object locus, from \citet{kauffmann03}. Representative error bars can be found in the upper left corner of each plot. All star forming regions, except for one, fall within the star forming locus in each of the three diagrams.}
    \label{fig:bpt}
\end{figure*}

Figure \ref{fig:bpt} displays the location of all identified regions from all the galaxies on three emission line diagnostic diagrams \citep{bpt}, along with the extreme starburst lines and the composite line from \cite{kewley06} and \cite{kauffmann03} boundaries. While \cite{2020MNRAS.494.4751M} find 56 star forming regions across the 29 galaxies, at the lower resolution we are only able to identify 35 distinct star forming regions. We list the locations of the 35 newly defined regions in Table \ref{table:regions}, which also includes the corresponding regions identified by \cite{2020MNRAS.494.4751M}.

\begin{table*}
%\centering
\caption{Aperture definitions of the star forming regions}\label{table:regions}
\setlength{\tabcolsep}{17 pt}
\begin{tabular}{ccccccc}
\hline\hline
Region & R.A. & Dec. & $a$ &  & $\theta$ & \cite{2020MNRAS.494.4751M} \\
I.D. & (deg) & (deg) & (arcsec) & $b/a$ & (deg) & Region\\
(1) & (2) & (3) & (4) & (5) & (6) & (7) \\
\hline
1.1 & 318.0262 & 11.3452 & 10.0 & 0.6 & 10 & 1.1, 1.2, 1.3\\
2.1 & 256.7246 & 32.1702 & 6.0 & 0.7 & 70 & 2.1 \\
3.1 & 247.2359 & 39.6095 & 6.0 & 1.0 & \dots & 3.1 \\
4.1 & 61.8478 & $-6.6862$ & 6.0 & 0.7 & 120 & 4.1 \\
5.1 & 166.1879 & 45.1565 & 6.0 & 1.0 & \dots & 5.1 \\
6.1 & 213.4954 & 43.8930 & 6.0 & 1.0 & \dots & 6.1 \\
7.1 & 154.8364 & 46.5501 & 7.0 & 0.6 & 70 & 7.1, 7.2, 7.3 \\
8.1 & 46.7492 & $-0.8116$ & 6.0 & 0.8 & 10 & 8.1 \\
9.1 & 213.3655 & 43.9138 & 7.0 & 0.6 & 110 & 9.1, 9.2, 9.3 \\
10.1 & 40.3040 & $-0.8770$ & 6.0 & 0.7 & 60 & 10.1, 10.2 \\
11.1 & 225.1996 & 48.6075 & 6.0 & 1.0 & \dots & 11.1 \\
12.1 & 115.2247 & 40.0697 & 6.0 & 0.8 & 170 & 12.1, 12.2, 12.3 \\
13.1 & 120.1982 & 46.6922 & 6.0 & 0.8 & 160 & 13.1\\
13.2 & 120.1999 & 46.6888 & 6.0 & 0.8 & 160 & 13.3 \\
14.1 & 220.1605 & 53.4781 & 6.0 & 1.0 & \dots & 14.1 \\
15.1 & 44.1697 & $-0.2460$ & 6.5 & 0.8 & 140 & 15.1 \\
16.1 & 118.1210 & 49.8397 & 6.0 & 1.0 & \dots & 16.1, 16.2 \\
17.1 & 118.8558 & 48.4385 & 6.0 & 1.0 & \dots & 17.1 \\
18.1 & 165.1179 & 44.2610 & 7.0 & 0.6 & 40 & 18.1 \\
19.1 & 119.9948 & 31.8934 & 6.0 & 1.0 & \dots & 19.1 \\
20.1 & 171.1020 & 23.6490 & 6.0 & 1.0 & \dots & 20.1\\
20.2 & 171.1057 & 23.6495 & 6.0 & 1.0 & \dots & 20.3 \\
21.1 & 248.1097 & 39.5177 & 6.0 & 1.0 & \dots & 21.1 \\
22.1 & 137.9836 & 27.8992 & 6.0 & 1.0 & \dots & 22.1 \\
23.1 & 206.4690 & 26.7726 & 6.0 & 1.0 & \dots & 23.2, 23.4 \\
23.2 & 206.4710 & 26.7770 & 6.0 & 1.0 & \dots & 23.1 \\
24.1 & 178.3380 & 52.3455 & 6.0 & 1.0 & \dots & 24.1, 24.2 \\
25.1 & 247.1632 & 40.1255 & 8.0 & 0.5 & 140 & 25.2, 25.3, 25.4\\
25.2 & 247.1626 & 40.1220 & 6.0 & 1.0 & \dots & 25.6\\
25.3 & 247.1675 & 40.1235 & 6.0 & 1.0 & \dots & 25.5\\
26.1 & 154.7685 & 46.4544 & 6.0 & 1.0 & \dots & 26.2, 26.3 \\
26.2 & 154.7715 & 46.4515 & 10.0 & 0.4 & 90 & 26.1, 26.5 \\
27.1 & 176.1242 & 55.5785 & 8.0 & 0.5 & 0 & 27.1\\
28.1 & 113.4005 & 45.9434 & 6.0 & 1.0 & \dots & 28.1 \\
29.1 & 46.7171 & $-0.8966$ & 6.0 & 1.0 & \dots & 29.1 \\
\hline
\end{tabular}
\\ Column 1: The ID associated with each region. Column 2: Right Ascension in degrees. Column 3: Declination in degrees. Column 4: Semimajor axis in arcseconds. Column 5: Ratio of semiminor to semimajor axis. Column 6: Angle of region with respect to vertical in degrees. Column 7: \cite{2020MNRAS.494.4751M} region I.D.(s) that is/are included in region. 
\end{table*}

\subsection{Properties of Star Forming Regions}
We calculated the strength of the 4000 \AA\ break, i.e., $\mathrm{D}_n(4000)$, for each region using maps in the SwiM VAC that were convolved and re-projected to the W3 band resolution \citep[SwiM VAC;][]{swim}. Here, we used the relation $\mathrm{D}_n(4000) = f_{\nu,\mathrm{red}}/f_{\nu,\mathrm{blue}}$, where $f_{\nu,\mathrm{red}}$ and $f_{\nu,\mathrm{blue}}$ refer to the average flux densities over the rest-frame wavelength ranges of $4000-4100$ \AA\ and $3850-3950$ \AA\ respectively. Uncertainties on D$_n(4000)$ were drawn from MaNGA uncertainty maps and include the systematic uncertainties described by \cite{westfall19}. \cite{2020MNRAS.494.4751M} caution against using the \cite{B16} attenuation law in regions with D$_n(4000)>1.3$. In later analysis, we flagged regions with values greater than $1.3$, and any composite regions that were made up of multiple regions previously identified by \cite{2020MNRAS.494.4751M} where \textit{any} of the initial regions had large D$_n(4000)$ values. We checked our new D$_n(4000)$ values against those found by \cite{2020MNRAS.494.4751M} for their regions, and found that, within each galaxy, our values were comparable.

We also calculated the gas-phase metallicity for each region using the [\ion{N}{ii}]/[\ion{O}{ii}] ratio as described by \cite{kewleydopita}. The [\ion{N}{ii}]/[\ion{O}{ii}] ratio is a useful metallicity diagnostic for regions with 12 + log(O/H) $> 8.5$, and has no dependence on ionization parameter \citep{kewleydopita}. It is also relatively insensitive to the presence of DIG \citep{zhang}. We used emission line maps provided in the SwiM VAC that were convolved and re-projected to the WISE W3 resolution. Errors in the gas-phase metallicity were estimated from the spread of curves seen in Figure 3 of \cite{kewleydopita} to be $\sim 0.1$, given that our minimum measured gas-phase metallicity was $8.59$. The estimated error was dominated by this uncertainty, as uncertainties in the MaNGA maps and the systematic uncertainty in emission line flux as described by \cite{belfiore19} were both small in comparison.

Both the D$_n(4000)$ and gas-phase metallicity measurements for individual regions are provided in Table \ref{table:sfrquants}.

\begin{table*}
%\centering
\caption{Derived quantities for star forming regions}\label{table:sfrquants}
\setlength{\tabcolsep}{17 pt}
\begin{tabular}{ccccccc}
\hline\hline
Region & Galactocentric Distance & Area & & & & \\
I.D. & (kpc) & (kpc$^2$) & IRX & $\beta$ & $12+\log([\mathrm{O/H}])$ & D$_n(4000)$ \\
(1) & (2) & (3) & (4) & (5) & (6) & (7) \\
\hline
1.1 & 0.6 & 31.1 & 0.21 $\pm$ 0.3 & $-$1.26 $\pm$ 0.2 & 8.59 $\pm$ 0.1 & 1.31 $\pm$ 0.06 \\
2.1 & 5.1 & 51.5 & 0.23 $\pm$ 0.3 & $-$1.09 $\pm$ 0.2 & 8.7 $\pm$ 0.1 & 1.23 $\pm$ 0.12 \\
3.1 & 0.9 & 123.8 & 0.66 $\pm$ 0.3 & $-$1.13 $\pm$ 0.2 & 8.94 $\pm$ 0.1 & 1.27 $\pm$ 0.05 \\
4.1 & 2.3 & 78.3 & 0.33 $\pm$ 0.3 & $-$1.29 $\pm$ 0.2 & 8.79 $\pm$ 0.1 & 1.3 $\pm$ 0.05 \\
5.1 & 0.4 & 23.6 & 0.76 $\pm$ 0.3 & $-$0.9 $\pm$ 0.2 & 9.05 $\pm$ 0.1 & 1.55 $\pm$ 0.06 \\
6.1 & 0.5 & 113.5 & 0.23 $\pm$ 0.3 & $-$1.15 $\pm$ 0.2 & 8.73 $\pm$ 0.1 & 1.33 $\pm$ 0.04 \\
7.1 & 0.4 & 48.2 & 0.37 $\pm$ 0.3 & $-$1.43 $\pm$ 0.2 & 8.83 $\pm$ 0.1 & 1.32 $\pm$ 0.03 \\
8.1 & 0.6 & 75.3 & 0.43 $\pm$ 0.3 & $-$1.07 $\pm$ 0.2 & 8.88 $\pm$ 0.1 & 1.28 $\pm$ 0.05 \\
9.1 & 0.8 & 152.0 & 0.28 $\pm$ 0.3 & $-$1.37 $\pm$ 0.2 & 8.82 $\pm$ 0.1 & 1.26 $\pm$ 0.04 \\
10.1 & 0.2 & 83.7 & 0.55 $\pm$ 0.3 & $-$1.03 $\pm$ 0.2 & 8.81 $\pm$ 0.1 & 1.26 $\pm$ 0.06 \\
11.1 & 0.4 & 126.7 & 0.91 $\pm$ 0.3 & $-$0.7 $\pm$ 0.2 & 8.97 $\pm$ 0.1 & 1.4 $\pm$ 0.07 \\
12.1 & 0.8 & 151.3 & 0.13 $\pm$ 0.3 & $-$1.43 $\pm$ 0.2 & 8.66 $\pm$ 0.1 & 1.13 $\pm$ 0.01 \\
13.1 & 5.7 & 31.7 & 0.46 $\pm$ 0.3 & $-$1.26 $\pm$ 0.2 & 8.79 $\pm$ 0.1 & 1.21 $\pm$ 0.02 \\
13.2 & 5.3 & 31.7 & 0.61 $\pm$ 0.3 & $-$1.29 $\pm$ 0.2 & 8.73 $\pm$ 0.1 & 1.23 $\pm$ 0.02 \\
14.1 & 0.2 & 88.1 & 1.37 $\pm$ 0.3 & $-$0.48 $\pm$ 0.2 & 8.9 $\pm$ 0.1 & 1.46 $\pm$ 0.07 \\
15.1 & 1.3 & 31.2 & 0.54 $\pm$ 0.3 & $-$1.3 $\pm$ 0.2 & 8.89 $\pm$ 0.1 & 1.22 $\pm$ 0.01 \\
16.1 & 5.4 & 55.9 & 0.92 $\pm$ 0.3 & $-$0.97 $\pm$ 0.2 & 8.98 $\pm$ 0.1 & 1.47 $\pm$ 0.07 \\
17.1 & 0.2 & 33.4 & 1.0 $\pm$ 0.3 & $-$1.03 $\pm$ 0.2 & 9.09 $\pm$ 0.1 & 1.46 $\pm$ 0.03 \\
18.1 & 0.6 & 74.5 & 1.66 $\pm$ 0.3 & $-$0.96 $\pm$ 0.2 & 8.94 $\pm$ 0.1 & 1.45 $\pm$ 0.05 \\
19.1 & 0.3 & 173.3 & 1.03 $\pm$ 0.3 & $-$0.67 $\pm$ 0.2 & 9.05 $\pm$ 0.1 & 1.28 $\pm$ 0.03 \\
20.1 & 8.0 & 48.4 & 0.53 $\pm$ 0.3 & $-$1.42 $\pm$ 0.2 & 8.97 $\pm$ 0.1 & 1.33 $\pm$ 0.03 \\
20.2 & 8.5 & 48.4 & 0.65 $\pm$ 0.3 & $-$1.14 $\pm$ 0.2 & 8.97 $\pm$ 0.1 & 1.33 $\pm$ 0.03 \\
21.1 & 0.5 & 63.4 & 0.67 $\pm$ 0.3 & $-$1.25 $\pm$ 0.2 & 9.11 $\pm$ 0.1 & 1.59 $\pm$ 0.04 \\
22.1 & 1.2 & 226.5 & 0.88 $\pm$ 0.3 & $-$0.68 $\pm$ 0.2 & 9.01 $\pm$ 0.1 & 1.3 $\pm$ 0.03 \\
23.1 & 13.7 & 81.8 & 0.54 $\pm$ 0.3 & $-$1.37 $\pm$ 0.2 & 8.97 $\pm$ 0.1 & 1.26 $\pm$ 0.03 \\
23.2 & 8.4 & 81.8 & 0.8 $\pm$ 0.3 & $-$1.16 $\pm$ 0.2 & 9.13 $\pm$ 0.1 & 1.41 $\pm$ 0.03 \\
24.1 & 3.7 & 133.1 & 0.76 $\pm$ 0.3 & $-$0.79 $\pm$ 0.2 & 9.04 $\pm$ 0.1 & 1.27 $\pm$ 0.03 \\
25.1 & 8.6 & 42.0 & 0.86 $\pm$ 0.3 & $-$1.36 $\pm$ 0.2 & 8.93 $\pm$ 0.1 & 1.31 $\pm$ 0.02 \\
25.2 & 10.0 & 47.2 & 0.83 $\pm$ 0.3 & $-$1.33 $\pm$ 0.2 & 8.99 $\pm$ 0.1 & 1.3 $\pm$ 0.03 \\
25.3 & 11.6 & 47.2 & 0.84 $\pm$ 0.3 & $-$1.33 $\pm$ 0.2 & 8.99 $\pm$ 0.1 & 1.31 $\pm$ 0.03 \\
26.1$^a$ & 13.4 & 56.4 & \dots & \dots & 8.87 $\pm$ 0.1 & 1.2 $\pm$ 0.01 \\
26.2$^a$ & 12.1 & 62.7 & \dots & \dots & 8.91 $\pm$ 0.1 & 1.24 $\pm$ 0.02 \\
27.1 & 6.1 & 336.9 & 1.16 $\pm$ 0.3 & $-$0.22 $\pm$ 0.2 & 8.97 $\pm$ 0.1 & 1.72 $\pm$ 0.08 \\
28.1 & 3.1 & 242.5 & 1.29 $\pm$ 0.3 & $-$0.84 $\pm$ 0.2 & 9.0 $\pm$ 0.1 & 1.27 $\pm$ 0.01 \\
29.1 & 0.4 & 386.5 & 0.74 $\pm$ 0.3 & $-$1.38 $\pm$ 0.2 & 9.08 $\pm$ 0.1 & 1.4 $\pm$ 0.03 \\
\hline
\end{tabular}
\\ Column 1: Region I.D. Column 2: Distance of region from the nominal center of the galaxy in kiloparsecs. Column 3: Area of the region in kpc$^2$. Column 4: Calculated value of IRX. Column 5: Calculated value of $\beta$. Column 6: Gas-phase metallicity. We calculate this using the [\ion{N}{ii}]/[\ion{O}{ii}] ratio, as described by \cite{kewleydopita}. Column 7: The systematic uncertainty in $\mathrm{D}_n(4000)$ as reported by \cite{westfall19} is included in the reported error. \\
$^a$ We do not calculate values for galaxy 26 that rely on \textit{Swift}/UVOT measurements due to the supernova light that contaminates the image.
\end{table*}

\subsection{The IRX--$\beta$ Relation}
\subsubsection{Calculation of $\beta$}
We calculated $\beta$ following the same procedure as \cite{2020MNRAS.494.4751M}. First we simulated a 100 Myr continuous star formation continuum model using \texttt{Starburst99} \citep{starburst99}, with a star formation rate of $1\; \mathrm{M}_\odot\;\mathrm{yr}^{-1}$. The resulting SED was then redshifted to the frame of each galaxy and reddened to a range of $E(B-V)$ values using the \cite{B16} attenuation law. We then measured $\beta$ as a function of $E(B-V)$ for the simulated SED using the 10 continuum windows defined by \cite{Calzetti94}. 

We calculated the uvw2--uvw1 colour for each redshifted SED using the UVOT filter transmission curves. These curves, as well as an example redshifted and reddened SED can be seen in Figure 1 of \cite{2020MNRAS.494.4751M}. Notably, the uvw2 and uvw1 transmission curves fit along either side of 2175 \AA, allowing them to measure UV colour in a way that is not affected by the strength of the 2175 \AA\ bump in the underlying attenuation curve. Finally, we measured the uvw2$-$uvw1 colour for each of our identified star forming regions and inferred their $\beta$ values by using the results of the simulations described above. These $\beta$ values can be found in Table \ref{table:sfrquants}. We also converted our values of $\beta$ to $\beta_{\rm Galex}$, i.e., the values $\beta$ would have had if they were measured via the FUV and NUV filters of GALEX \citep{galex} using the relationship defined by \cite{2020MNRAS.494.4751M}. See Section 4.3 of \cite{2020MNRAS.494.4751M} for further discussion of this conversion and its necessity. The error associated with this conversion represents the largest source of uncertainty in the calculation of $\beta$, and dominates the size of the error bars.

\subsubsection{Calculation of IRX}
\begin{figure*}
\centering
\includegraphics[width=0.45\textwidth]{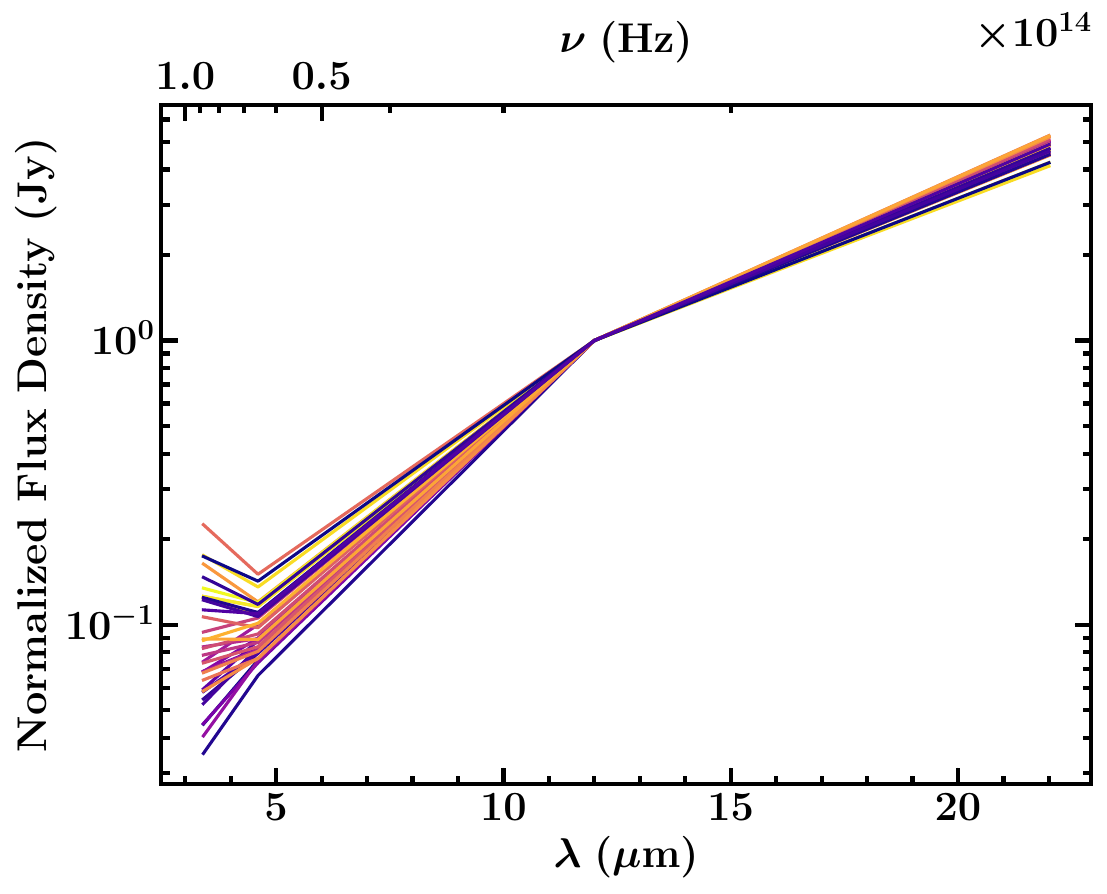}
\hskip 1cm
\includegraphics[width=0.45\textwidth]{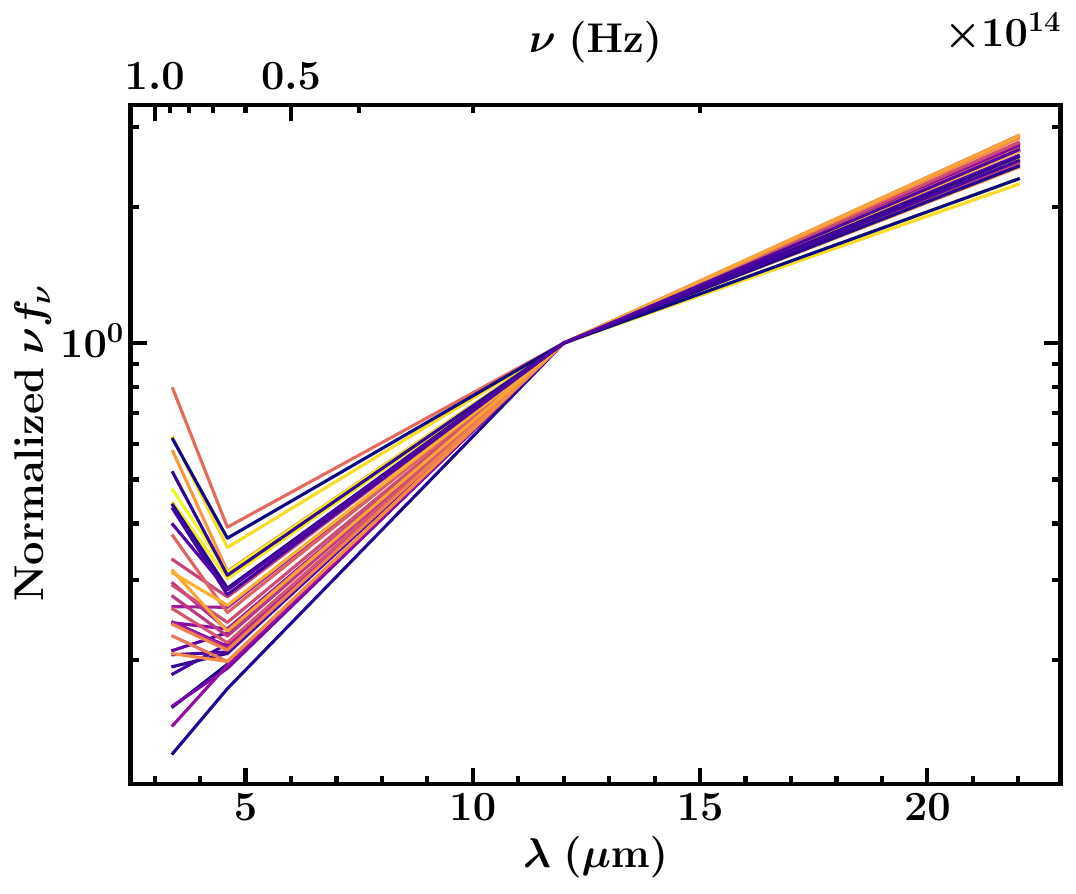}
\caption{We plot flux values in the star forming regions in each of the WISE bands. On the left is the normalized flux density plotted against the wavelength, and on the right is normalized $\nu$F$_\nu$ against wavelength.  We normalize each value by the flux in the W3 band for each region. We do this to test the validity of using a relationship between $L_{12}$ and $L_{\mathrm{TIR}}$ for our regions. We see that, generally, the shapes are similar and appear to diverge mostly at short wavelength, but that most of the flux is in the portions of the SED at longer wavelength, where W3 and W4 are located.}
\label{fig:normsed}
\end{figure*}

The IRX is defined as the log of the ratio between the total IR luminosity and the FUV luminosity \citep{meurer99}. Here, we used the IR luminosity in the W3 band to determine the total IR luminosity, following the scaling relation defined by \cite{cluver}. This relationship connects the monochromatic luminosity at 12~$\mu$m to the total IR luminosity via $\log \mathrm{L}_\mathrm{TIR} = 0.889\log \mathrm{L}_{12\mu \mathrm{m}} + 2.21$. \cite{cluver} found that this relation predicts L$_\mathrm{TIR}$ well in normal galaxies. This conversion represented the largest single source in error in the calculation of the IRX for each region, with an uncertainty of $\sim 0.2$ dex. Additional uncertainties in flux measurements accounted for the additional $\sim 0.1$ dex in the error calculation. To verify the applicability of the relation from \cite{cluver}, we measured flux values for each region in each of the WISE bands (W1, W2, W3, and W4). We then normalized these by the flux in the W3 band and plotted the resulting SED\null. As shown in Figure \ref{fig:normsed}, in both the plot of normalized flux density against frequency and $\nu$F$_\nu$ against frequency, emission at frequencies lower (and thus wavelengths longer) than the W3 band dominates. While the SED shape diverges primarily at higher frequencies, the bulk of the observed flux is emitted at lower frequencies, probed by W3 and W4. Thus, we concluded that the \cite{cluver} relation is sufficient to estimate L$_\mathrm{TIR}$ for our analysis.

We also calculated the FUV luminosity density at 1600~\AA\ for each star forming region. Since UVOT only has NUV filters, we translated the flux from the filter with the shortest wavelength, uvw2 ($\lambda_{\mathrm{cent}}=1928$~\AA\null) to that at 1600~\AA. To convert between the fluxes at 1928~\AA\ and 1600~\AA, we assumed that the two fluxes were related through: $F_{1600}=F_{1928}(1600/1928)^\beta$, effectively assuming that the observed UV continuum follows a power law of the form $F(\lambda)\propto\lambda^\beta$ in the wavelength range $1250\leq\lambda\leq2650$ \AA\  \citep{Calzetti94}. We used the calculated $\beta$ value for each region to scale our uvw2 flux measurements appropriately.

The calculated total IR luminosity and the luminosity density at $1600$ \AA\ were then used to find IRX values for each region following $\mathrm{IRX}\equiv \log (L_{\mathrm{TIR}}/L_{\mathrm{FUV}})$ \citep{meurer99}. IRX values for each region can be found in Table \ref{table:sfrquants}.

\subsubsection{Fitting the IRX--$\beta$ Relation}
\begin{figure}
    \centering
    \includegraphics[width=0.95\linewidth]{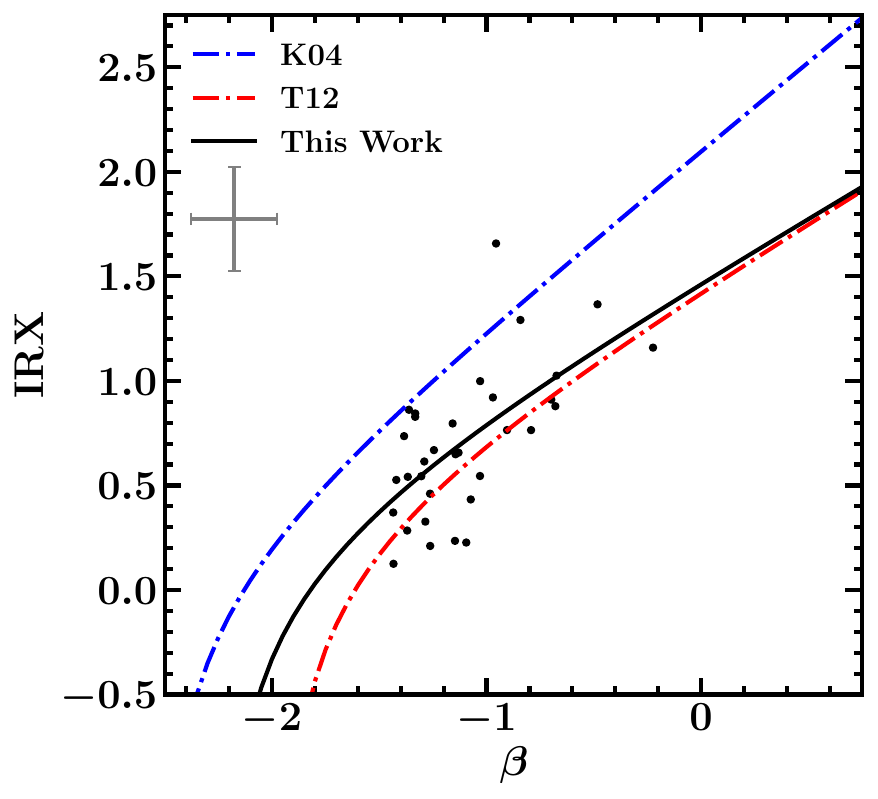}
    \caption{Star-forming regions in IRX-$\beta$ space compared to relationships for starburst galaxies \citep[blue dashed line;][]{kong04} and local galaxies from \citet[red dot dashed line;][]{takeuchi12}. Additionally, we plot our best fit relationship given in equation \ref{eq:irxb_reg} (solid black line). Representative error bars can be found in the upper left corner. We find that our star forming regions follow a relationship similar to that followed by the local galaxies sampled by \citet{takeuchi12}.}
    \label{fig:irxb}
\end{figure}

We combined our measured values for IRX and $\beta$ for each of the regions to examine the relationship between the two. We show the resulting IRX-$\beta$ relation and its best fit in Figure \ref{fig:irxb}. The relation was fit using non-linear least squares regression to the functional form given by \cite{meurer99}, allowing all three free parameters to be free to vary simultaneously. We obtained
\begin{equation}\label{eq:irxb_reg}
        \mathrm{IRX} = \log[10^{0.4(1.51\beta_{\rm{reg}}+3.33)}-1] + 0.15\; ,
\end{equation}
where $\beta_{\rm{reg}}$ is the NUV slope for that region (the free parameters are $a=1.51$,  $b=3.33$, and $c=0.15$).

Figure \ref{fig:irxb} also shows other IRX-$\beta$ relationships for local and starburst galaxies \citep{kong04, takeuchi12}. We note that our best-fitting relationship is similar to the one determined by \cite{takeuchi12} for the integrated light from local galaxies. Our points have a large scatter about the best fitting relation, but seem to follow expected relationships for integrated galaxy light. The scatter of the star-forming regions about the relationship, however, is greater than can be attributed to the size of the error bars alone. The scatter is similar to that shown in other studies of this kind, both on galactic \citep[e.g.][]{kong04, takeuchi12, nagaraj} and subgalactic scales \citep[e.g.][]{boquien}.

\begin{figure}
    \centering
    \includegraphics[width=0.95\linewidth]{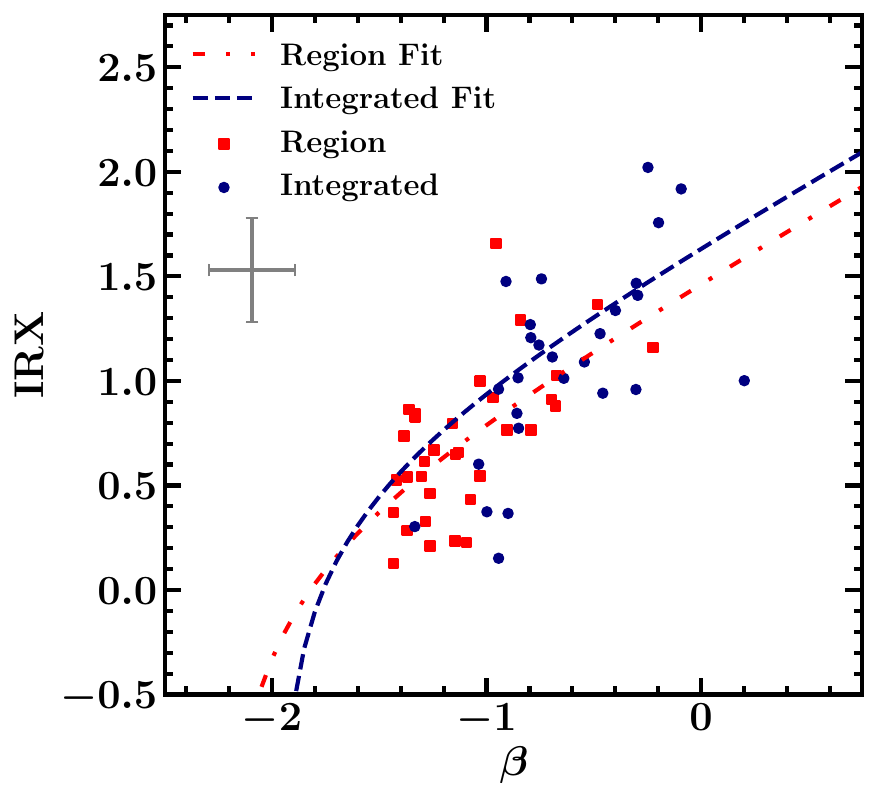}
    \caption{The IRX--$\beta$ relation for star forming regions (red squares and dot dashed line) and for integrated galaxy light (blue circles and dashed line). Best fits found through non-linear least squares regression are noticeably similar for both the individual regions (see equation \ref{eq:irxb_reg}) and integrated galaxy light (see equation \ref{eq:irxb_int}). The representative error bar for each set of points is plotted in the upper left, and is the same for both integrated galaxy light and light from individual regions. This is because the dominant source of error in the IRX arises from the conversion between W3 luminosity and L$_\mathrm{TIR}$ and the dominant source of error in $\beta$ arises from conversion to $\beta_\mathrm{GALEX}$. These sources of error are present and dominant in the measurements for the galaxies as a whole and the individual regions. We find that star forming regions and integrated galaxy light from their hosts tend to follow very similar IRX--$\beta$ relations.}
    \label{fig:irxb_int}
\end{figure}

In Figure \ref{fig:irxb_int}, we compare the IRX-$\beta$ relationship for the integrated light of our sample of galaxies to that for the individual star forming regions. We fitted the integrated galaxy light data following the same procedure as for the individual regions to obtain a relationship of the same functional form as Equation \ref{eq:irxb_reg}, but for the NUV slope for integrated galaxy light, namely
\begin{equation}\label{eq:irxb_int}
    \mathrm{IRX} = \log[10^{0.4(1.473\beta_{\rm{int}}+2.889)}-1] + 0.507.
\end{equation}
(the free parameters are $a=1.473$, $b=2.889$, and $c=0.507$).

We find that individual regions and integrated galaxy light seem to follow very similar IRX-$\beta$ relations. This effect may be a result of the size of our regions -- due to the resolution of the WISE W3 band, our minimum aperture size is fairly large and our regions can encompass significant portions of their host galaxies.

\subsubsection{Effect of S/N on $\beta$ and IRX}
All observed galaxies had a minimum S/N of 15 in the integrated galaxy light for both \textit{Swift}/UVOT filters used, and similarly a minimum S/N of 10 in the WISE W3 band. If we were to apply a more stringent cut on the S/N of the galaxies, neither the level of scatter seen in IRX and $\beta$ nor the overall best fit relation would change significantly. This means that the scatter we see is not a result of our choice in S/N but is likely intrinsic. When we investigate the effect of stricter S/N requirements on other properties of the regions, we tend to find a marginally higher average SFR, although the average gas-phase metallicity of the regions does not change. Hence, we conclude that imposing more stringent S/N cuts does not change our conclusions.

\section{Dependence of the IRX--$\beta$ Relation on Physical Parameters}\label{sec:analysis}
\begin{figure*}
\hbox{
\hbox to 3.5truein{\includegraphics[height=2.5truein]{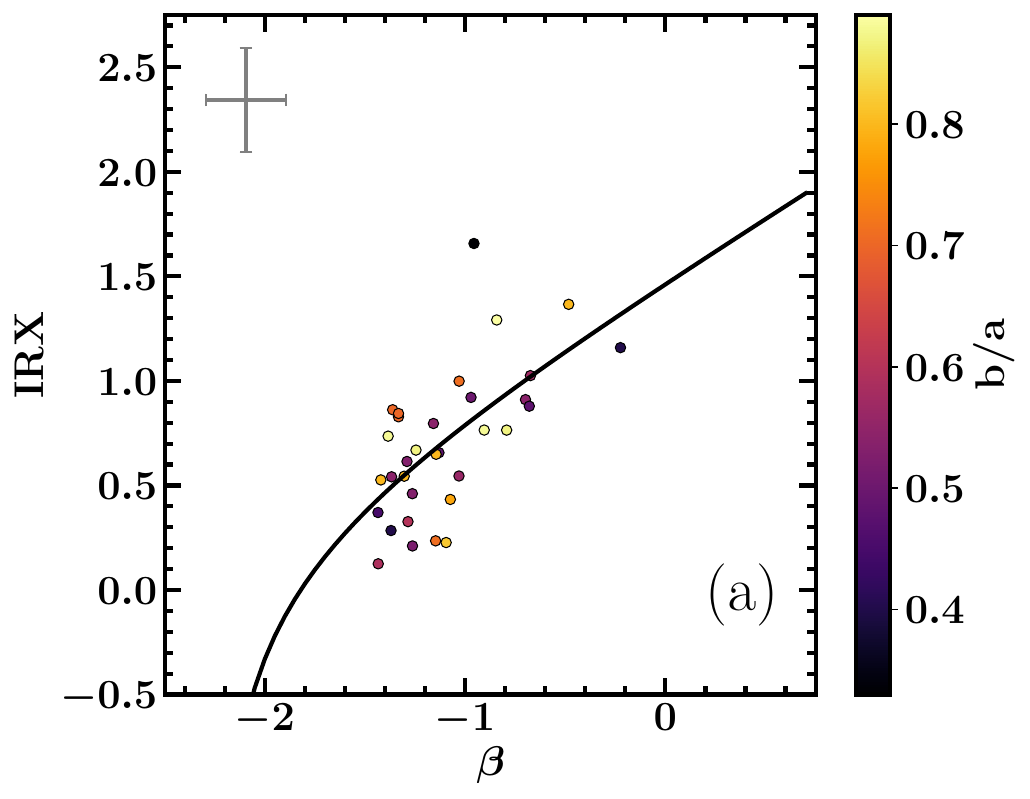}\hfill}
\hbox to 3truein{\includegraphics[height=2.5truein]{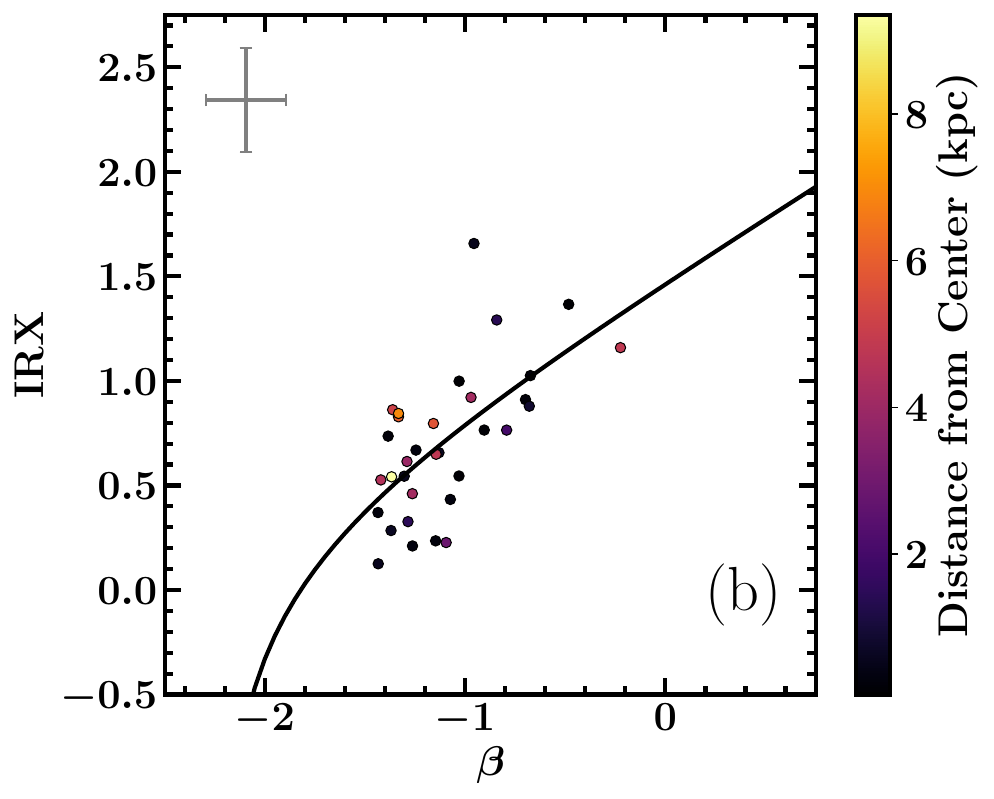}}
}
\vskip 0.1truein
\hbox{
\hbox to 3.5truein{\includegraphics[height=2.5truein]{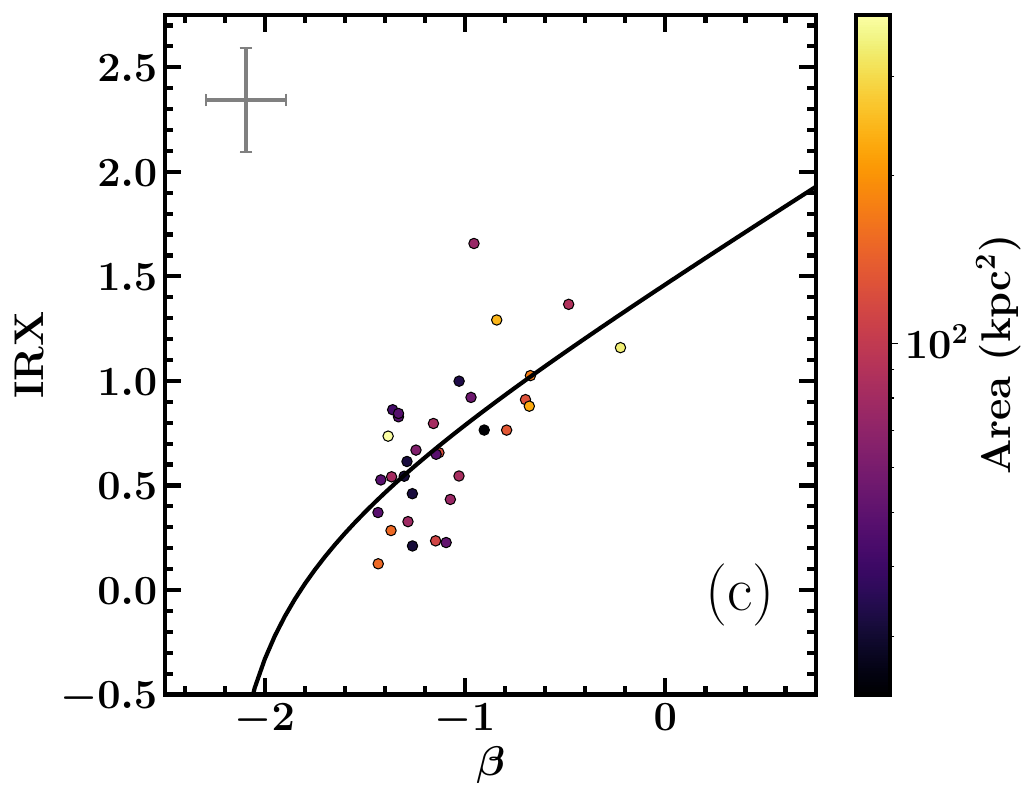}\hfill}
\hbox to 3truein{\includegraphics[height=2.5truein]{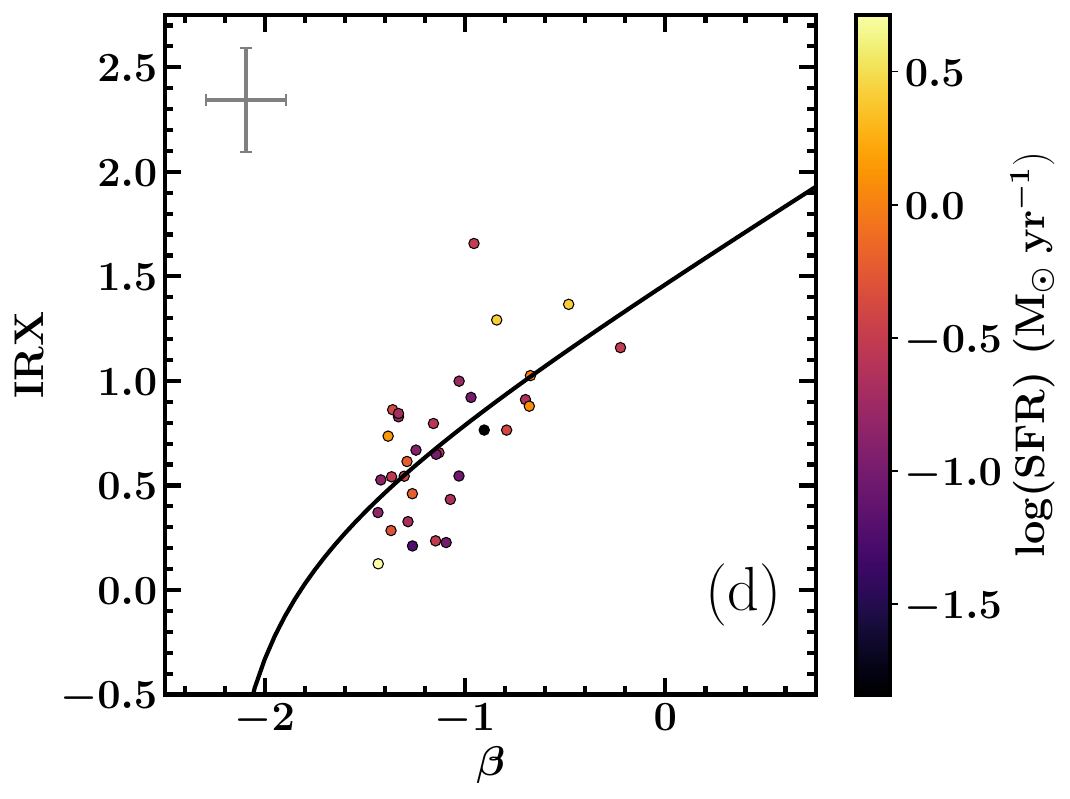}}
}
\vskip 0.1truein
\hbox{
\hbox to 3.5truein{\includegraphics[height=2.5truein]{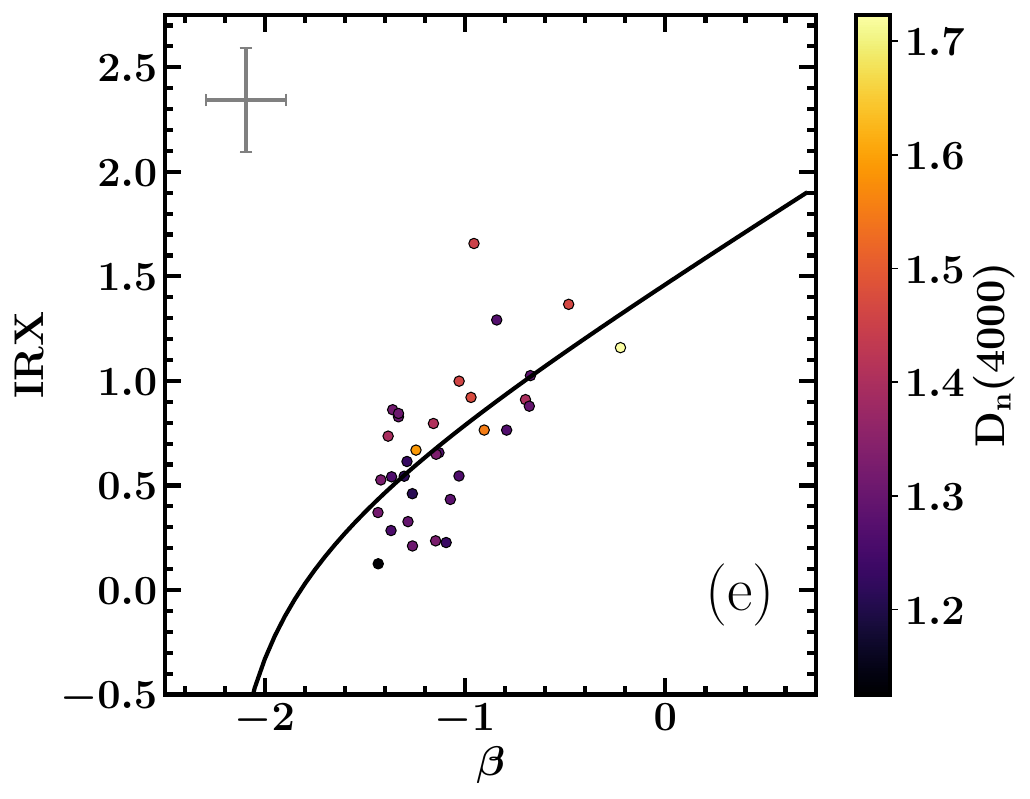}\hfill}
\hbox to 3truein{\includegraphics[height=2.5truein]{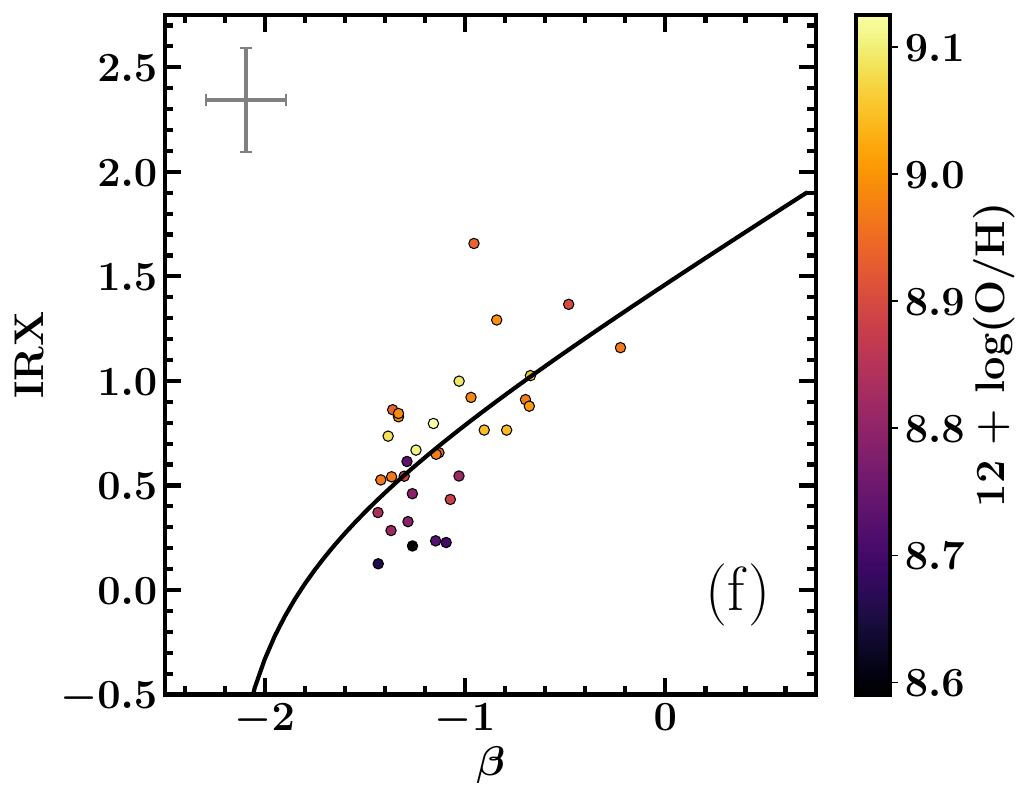}}
}
\caption{IRX-$\beta$ points colored by a number of observable parameters. We check correlations between (a) inclination, measured by $b/a$, (b) galactocentric distance in kpc, (c) area of the region in kpc$^2$, (d) the star formation rate, (e) the average stellar age of the region, measured by D$_n(4000)$, and (f) the gas-phase metallicity of the region. The characteristic error bars for the regions are shown in the upper left corner. We find the most convincing correlations exist between IRX-$\beta$ and D$_n(4000)$ and IRX-$\beta$ and gas-phase metallicity.}
\label{fig:correlations}
\end{figure*}

We investigated the relationship between IRX and $\beta$ for our star forming regions by comparing the results with a number of other properties, including galaxy inclination, measured by taking the ratio of the observed minor and major axes, and the galactocentric radius, area, SFR, age, as measured by D$_n(4000)$, and gas-phase metallicity of each region. We show colour-coded plots to illustrate the effect of each of these properties in Figure \ref{fig:correlations}.

We investigated the effect of inclination on the IRX-$\beta$ relationship in Figure \ref{fig:correlations}(a).  We found that while there is no overall relationship between the inclination of the galaxy and the IRX-$\beta$ relationship, inclination may explain the location of region 18.1. This region has the largest IRX value of our sample, and lies furthest from the locus of points in both Figures \ref{fig:irxb} and \ref{fig:correlations}(a) Its host galaxy is also the most inclined of our sample, with $b/a = 0.33$. \cite{salimboquien19} note that galaxies with $b/a$ values less than $0.3$ tend to fall significantly above the IRX-$\beta$ relation and exhibit more scatter than galaxies that are face-on. While this region has a value slightly above the threshold listed by \cite{salimboquien19}, it is sufficiently close to that limit to explain the increased scatter from the best-fit relation. Aside from this object, however, a Spearman rank order test showed no evidence for a correlation of IRX or $\beta$ with inclination, with p-values, which represent the probability that the observed correlation is due to chance, of 0.79 and 0.97 respectively. We also note that our sample seems to be biased towards galaxies that are face-on. Further analysis on a more balanced sample of inclinations would be needed to confirm this lack of relation. 

Because dust attenuation is known to decrease with increasing galactocentric radius, we explored the effect of position within a galaxy on a star forming region's location in the IRX-$\beta$ diagram, and show the results in Figure \ref{fig:correlations}(b). IRX and galactocentric distance have a Spearman rank order test p-value of 0.82, meaning that there is no statistically significant correlation between the two. We find that $\beta$ and galactocentric distance have a Spearman correlation coefficient of -0.32 and a p-value of 0.07, which is nearly, but not quite 2$\sigma$ significance. This potential negative correlation implies that as galactocentric distance increases, $\beta$ becomes more negative.

Given the lower resolution of our final images, some newly defined regions are no longer kpc-sized, but are rather on the order of 10~kpc across. We therefore plotted the relationship between the size of a region and its IRX and $\beta$ values to see what effect, if any, the larger apertures have on the measured properties. The results are shown in Figure \ref{fig:correlations}(c). When we performed statistical analysis, we found that there is no convincing trend between the area of a region and its value of IRX or $\beta$. The associated p-values from the Spearman rank order test were 0.28 and 0.59 respectively.

The effect of aperture size on the relation can also be explored by comparing the IRX-$\beta$ relation found for individual star-forming regions with the relation found for integrated galaxy light. We compared these relations in Figure \ref{fig:irxb_int}. We see in that figure that IRX-$\beta$ measurements for integrated galaxy light follow a very similar relationship to the light from individual star forming regions. We also see that integrated galaxy light tends to have both more positive $\beta$ values and higher IRX\null. The median values of IRX and $\beta$ for the integrated galaxy light are 1.05 and $-0.74$ respectively, whereas those values for the regions are 0.59 and $-1.29$. This represents a statistically significant difference between the values measured for individual regions and those measured for the galaxy as a whole. This finding is reasonable given that older stellar populations, increased dust content, or a combination thereof could create redder IRX and $\beta$ measurements in the integrated galaxy light. 

We calculated the star formation rate of each region, and plotted it with IRX and $\beta$ to see if a convincing relationship existed between SFR and IRX-$\beta$ in Figure \ref{fig:correlations}(d). The SFR was measured using the relationship for de-reddened H$\alpha$ luminosity as described by \cite{kennicuttevans}. We do not find statistically significant relationships between SFR and IRX or $\beta$. The p-values measured for these potential correlations are 0.20 and 0.85 respectively.

\begin{figure*}
\centering
\includegraphics[width=0.47\textwidth]{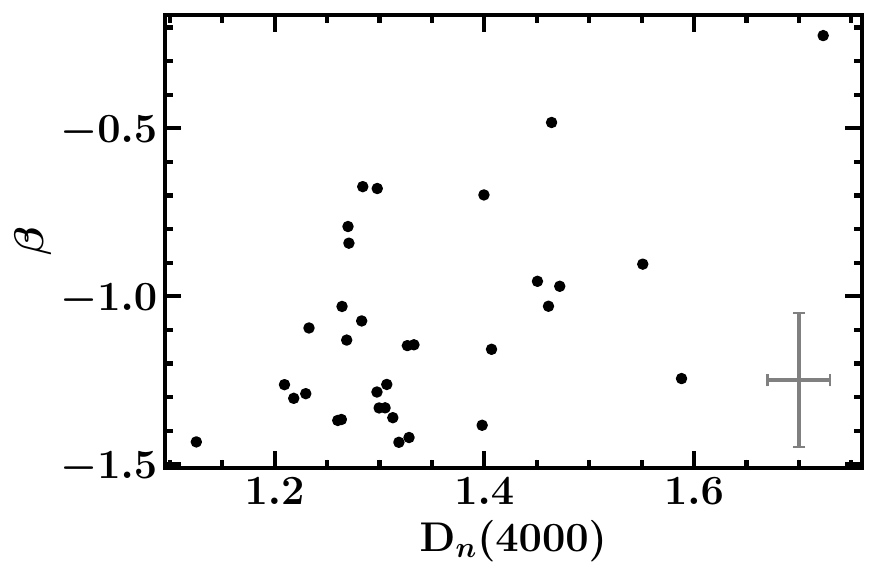}
\hskip 1cm
\includegraphics[width=0.45\textwidth]{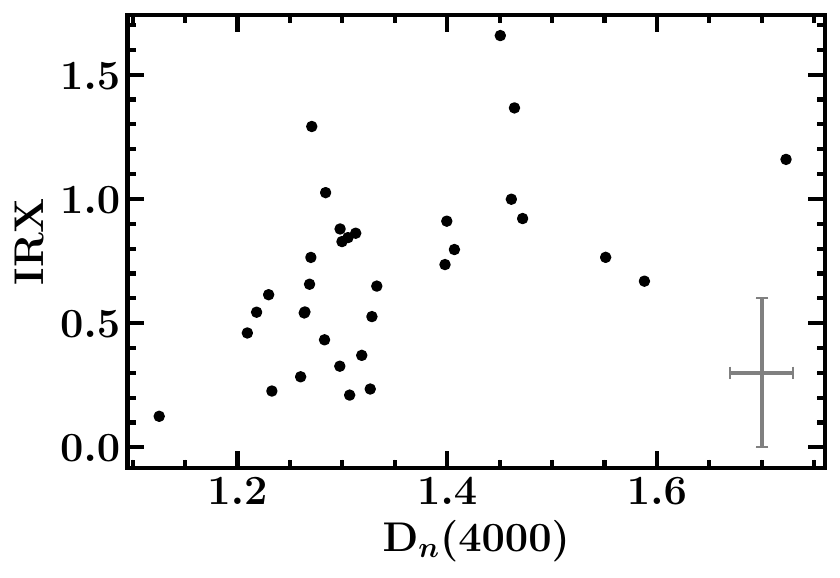}
\caption{We show the relationships between D$_n(4000)$ and $\beta$ (left) and D$_n(4000)$ and IRX (right). The average uncertainties in each figure are represented by the error bars in the upper left of each plot. We note a fairly strong correlation between IRX and D$_n(4000)$, with a  Spearman correlation coefficient of 0.54 and a p-value of 0.001, and a weaker correlation between $\beta$ and D$_n(4000)$ with a Spearman correlation coefficient of 0.35 and a p-value of 0.04.}
\label{fig:irxbd4000}
\end{figure*}

The amplitude of the 4000~\AA\ break can be used as a proxy for stellar age \citep{balogh}. Generally as light from older stars becomes more prominent, implying an increased average stellar age, D$_n(4000)$ also increases. We colour the points in our IRX-$\beta$ diagram according to their calculated D$_n(4000)$ values in Figure \ref{fig:correlations}(e). When we calculate correlation coefficients for IRX and D$_n(4000)$, and $\beta$ and D$_n(4000)$, we find that IRX and D$_n(4000)$ have a moderate positive correlation, with a Spearman correlation coefficient of 0.54, and a p-value of 0.001. We also find a slightly weaker positive correlation between D$_n(4000)$ and $\beta$. These two variables have a Spearman correlation coefficient of 0.35 and a p-value of 0.04. This means that IRX and D$_n(4000)$ are statistically correlated at the 99.9\% confidence level, and $\beta$ and D$_n(4000)$ are correlated at the 96\% confidence level. We show the relationship between D$_n(4000)$ and $\beta$ and D$_n(4000)$ and IRX in Figure \ref{fig:irxbd4000}. We also investigate the relationship between IRX-$\beta$ and another stellar age indicator, the H$\delta_A$ Lick Index \citep{worthey98}. We use convolved and reprojected SwiM VAC flux and continuum maps to calculate this index for each of our regions, following the equation defined in Section 5.4 of \cite{swim}. Similarly to D$_n(4000)$, we find a strong correlation between IRX and that index, showing that older stellar populations have higher values of IRX.

The positive correlations between the 4000~\AA\ break and both IRX and $\beta$ imply that older stellar populations tend to look redder in the IRX-$\beta$ space, and also might follow steeper attenuation curves. This is reasonable given that in regions with older average stellar populations, light from evolved stars becomes more prominent relative to the emission at 1600 \AA, which tends to probe younger stellar populations. This is also consistent with the general result from \cite{2020MNRAS.494.4751M}, who found that older regions tended to lie above the expected $\beta$--$\tau^{l}_B$ relationship.

We further investigated the difference between regions with D$_n(4000)>1.3$ and those with values $<1.3$, as \cite{2020MNRAS.494.4751M} found that those regions behave differently in $\tau^l_B$--$\beta$ space. In the IRX--$\beta$ relation, we do not find a significant difference in behavior between these two populations. Median $\beta$ values for the two populations are essentially the same and median IRX values for the two populations are separated by 0.15, where the group with D$_n(4000)>1.3$ has a higher IRX value. This difference, however, is smaller than the size of the individual error bar.

\begin{figure*}
\centering
\includegraphics[width=0.48\textwidth]{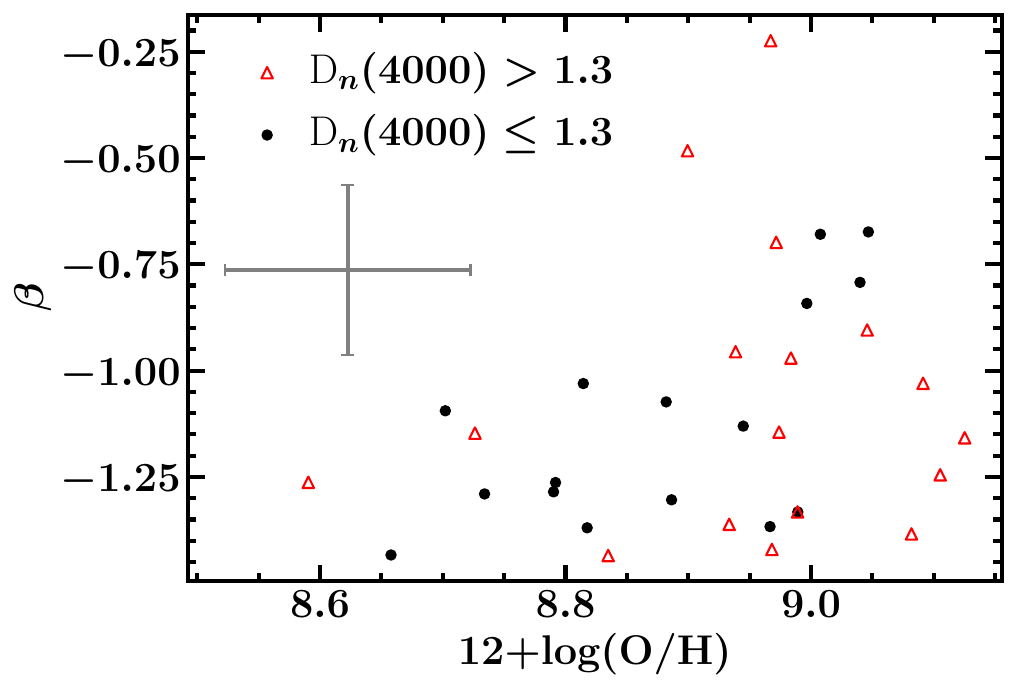}
\hskip 1cm
\includegraphics[width=0.45\textwidth]{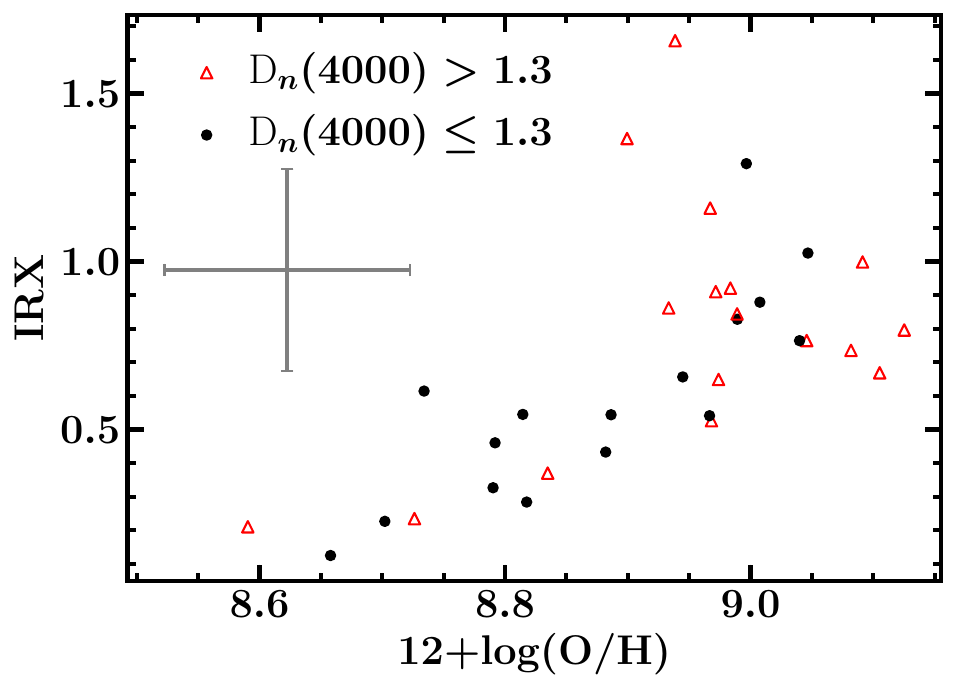}
\caption{We show the relationships between metallicity and $\beta$ (left) and metallicity and IRX (right), split by stellar age. Regions with D$_n(4000)>1.3$ are shown as red triangles and regions with D$_n(4000)\leq1.3$ are shown as black circles. The systematic uncertainties in each figure are represented by the error bars in the upper left of each plot. When all regions are included, we do not find a statistically significant correlation between metallicity and $\beta$, but do find a strong positive correlation between metallicity and IRX. When old regions, with D$_n(4000)>1.3$ are excluded, the scatter in both plots is noticeably smaller. We also recover strong positive correlations between IRX and metallicity and $\beta$ and metallicity when the older populations are excluded.}
\label{fig:irxmetsplit}
\end{figure*}

Finally, we explored the effect of gas-phase metallicity on the IRX-$\beta$ relationship. We calculated gas-phase metallicity using the [\ion{N}{ii}]/[\ion{O}{ii}] relationship defined by \cite{kewleydopita}, and corrected for internal reddening using the Balmer decrement. The plot of metallicity as a function of IRX-$\beta$ is shown in Figure \ref{fig:correlations}(f). We found that, while $\beta$ and metallicity do not have a statistically significant relationship, and have a p-value for correlation of 0.08, IRX and metallicity are strongly positively correlated, with a Spearman correlation coefficient of 0.67 and a p-value of 2.07$\times10^{-5}$. These two relationships are plotted in Figure \ref{fig:irxmetsplit}.

We also investigated the potential influence of aging stellar populations on our observed correlation between IRX and metallicity. To disentangle this effect, we separate the regions into those with D$_n(4000)\leq 1.3$ and $D_n(4000)>1.3$ as before, and measure correlations between the populations and IRX, $\beta$ and metallicity. When older stellar populations, those with $D_n(4000)>1.3$, are excluded, we find that the strength of the positive IRX-metallicity correlation is increased, with a Spearman correlation coefficient of 0.82 and a p-value of 9.84$\times10^{-5}$. Notably, we also recover a statistically significant, strong positive correlation between $\beta$ and metallicity, with a Spearman correlation coefficient of 0.55 and a p-value of 0.028. We show this effect in Figure \ref{fig:irxmetsplit}, where older populations are shown as red triangles and younger populations are shown as black circles. The scatter in the correlations between IRX and metallicity and $\beta$ and metallicity is noticeably smaller when older populations are excluded.

The correlation between metallicity and the IRX strongly implies that the shape of the attenuation law for a region is influenced by the metallicity of that region.  This result is consistent with the correlation between metallicity and IRX found by \cite{shivaei} in their sample of $z\sim 2-2.5$ galaxies, as well as the result found by \cite{boquien09} for sub-kpc sized star-forming regions within eight star-forming galaxies. The metallicity dependence thus appears to exist independent of aperture size. This result implies that the attenuation curve is steeper for galaxies with higher metallicities, whereas lower metallicity regions have grayer attenuation curves. Physically, this suggests that galaxies with lower metallicity have different dust grain compositions and thus attenuation curves than those with higher metallicity.

\section{Discussion and Conclusions}\label{sec:conc}
We examined the IRX--$\beta$ relation using a sample of previously identified star forming regions in 29 galaxies with both MaNGA and \textit{Swift}/UVOT observations \citep{2020MNRAS.494.4751M}. The sample is a subset of a larger catalogue that was created to investigate the quenching of star formation in nearby galaxies \citep{swim}. By supplementing the MaNGA and Swift/UVOT data with WISE W3 measurements, we were able to compare the IRX with the UV spectral slope for both the entire galaxies and kpc-sized star forming regions within the galaxies.

Integrated galaxy light follows the same IRX-$\beta$ relationship as the light from resolved star-forming regions, although the integrated light has both higher values of IRX and more positive values of $\beta$.  As a result, the curves that describe the IRX--$\beta$ relationship for the individual regions and the integrated galaxy light are very similar. This similarity may be taken to suggest that the IRX-$\beta$ relation that we observe in integrated galaxy light is driven by the contributions from star-forming regions to that light. Such a conclusion, however, is complicated by the relatively large minimum aperture size of our regions. The relationship we observe may be ultimately connected to the size of the apertures for a variety of reasons. The observed shift along the IRX-$\beta$ relation between the regions and the integrated light may also be caused by a combination of effects, such as aging stellar populations in the galaxy outside of the regions and increased dust content within the regions. In addition, previous studies have found significant variations in attenuation curves at different radii within galaxies \citep[e.g.][]{li20}. Our observed trend may be caused by some combination of changing attenuation and optical depths that, when seen together, mimics one attenuation curve. The large size of our star-forming regions might average over those radial effects, creating the impression that total galactic attenuation is driven by star-forming regions. This means that, in normal star-forming galaxies, kiloparsec scale star-forming regions are large enough to average sub-galactic processes into galactic-wide effects.

A number of properties, such as gas-phase metallicity \citep[e.g.][]{boquien09, shivaei}, stellar population age \citep[e.g.,][]{kong04, cortese08}, variations in star formation history \citep[e.g.,][]{kong04, narayanan18}, or variations in the underlying attenuation curve \citep[e.g.][]{narayanan18, salimboquien19}, have been invoked to explain the scatter in the IRX-$\beta$ relationship. While we have a relatively small sample of 29 galaxies and 35 star forming regions, we explore potential relationships between the IRX-$\beta$ relation and a number of parameters. We find that the distance of the regions from the centre of their host galaxies, the area of the region, and the inclination of the galaxy have little effect on where points fall in IRX-$\beta$ space, although we do note that the region with the largest IRX value has the smallest $b/a$ value. \cite{salimboquien19} note that edge-on galaxies tend to have higher IRX values and exhibit more scatter in the relationship than their face on counterparts. We attribute the placement of the aforementioned galaxy to this effect. We also note that our sample is largely made up of face-on galaxies. This selection bias might also play a role in the lack of relationship between IRX--$\beta$ and $b/a$ that we see here.

We find no statistically significant relationship between star formation rate and $\beta$ or IRX\null. Conversely, there is a moderate correlation between the value of D$_n(4000)$ and the IRX, where higher values of D$_n(4000)$ and thus older stellar populations appear to have higher IRX values. We also find a weaker, but still significant, correlation between $\beta$ and D$_n(4000)$. In contrast, we do not find any significant difference in the IRX-$\beta$ relationship between populations of regions with values of D$_n(4000)$ greater or less than 1.3, unlike the trend found by \cite{2020MNRAS.494.4751M}. We do, however, see a difference in the behavior of these two populations in the IRX-metallicity and $\beta$ metallicity spaces.

Of the parameters we considered, we see the strongest correlation between IRX and the gas-phase metallicity, $12 + \log(\mathrm{O/H})$. This is a strong positive correlation, where higher values of IRX correlate with higher metallicities. When regions with high values of D$_n(4000)$ are excluded and only regions with younger average ages are considered, the correlation between IRX and metallicity becomes stronger, and a statistically significant relationship between $\beta$ and metallicity is recovered. This result is consistent with previous work both in sub kpc-sized star-forming regions and in samples of $z\sim2-2.5$ galaxies. Therefore, \textit{regardless} of the aperture used, metal-rich regions will have steeper attenuation curves, whereas lower metallicity regions have grayer (i.e., flatter) attenuation curves. Observationally, this means that galaxies with higher metallicities should look redder than their lower metallicity counterparts.

Finally, we found that the scatter in IRX-$\beta$ does not decrease with more stringent requirements in S/N. When viewed in conjunction with the lack of correlation between IRX-$\beta$ and most of the physical parameters we investigate, we conclude that the scatter in IRX-$\beta$ at kiloparsec scales may be driven by intrinsic physical differences between different regions. These differences could largely be related to the gas-phase metallicity and the average stellar age of the regions, although the galactocentric radius may also contribute.

While our results are intriguing, they are somewhat limited due to our relatively small sample size. In a forthcoming paper, the second data release of the SwiM VAC (Molina et al., in preparation) will expand the number of galaxies with both \textit{Swift}/UVOT and MaNGA data from 150 to 559. This expanded sample will provide a significant increase in the number of star-forming galaxies available, and can be used to verify the trends noted here.

\section*{Acknowledgements}
We thank the referee, Tsutomu T. Takeuchi, for his thorough reading of the manuscript and helpful comments.

This work was supported by the NASA ADAP program through grant 80NSSC20K0436. 

This work was supported by funding from Ford Foundation Postdoctoral Fellowship, administered by the National Academies of Sciences, Engineering, and Medicine, awarded to MM in 2021-2022. The work of MM is supported in part through a fellowship sponsored by the Willard L. Eccles Foundation.  

RY acknowledges support by the Hong Kong Global STEM Scholar scheme, by the Hong Kong Jockey Club Charities Trust through the JC STEM Lab of Astronomical Instrumentation, and a grant from the Research Grants Council of the Hong Kong Special Administrative Region, China [Project No: CUHK 14302522].

MB gratefully acknowledges support by the ANID BASAL project FB210003 and from the FONDECYT regular grant 1211000.

The Institute for Gravitation and the Cosmos is supported by the Eberly College of Science and the Office of the Senior Vice President for Research at the Pennsylvania State University.

Funding for the Sloan Digital Sky 
Survey IV has been provided by the 
Alfred P. Sloan Foundation, the U.S. 
Department of Energy Office of 
Science, and the Participating 
Institutions. 

SDSS-IV acknowledges support and 
resources from the Center for High 
Performance Computing  at the 
University of Utah. The SDSS-IV 
website is www.sdss4.org.

SDSS-IV is managed by the 
Astrophysical Research Consortium 
for the Participating Institutions 
of the SDSS Collaboration including 
the Brazilian Participation Group, 
the Carnegie Institution for Science, 
Carnegie Mellon University, Harvard-Smithsonian Center for 
Astrophysics, the Chilean Participation 
Group, the French Participation Group, 
Instituto de Astrof\'isica de 
Canarias, The Johns Hopkins 
University, Kavli Institute for the 
Physics and Mathematics of the 
Universe (IPMU) / University of 
Tokyo, the Korean Participation Group, 
Lawrence Berkeley National Laboratory, 
Leibniz Institut f\"ur Astrophysik 
Potsdam (AIP),  Max-Planck-Institut 
f\"ur Astronomie (MPIA Heidelberg), 
Max-Planck-Institut f\"ur 
Astrophysik (MPA Garching), 
Max-Planck-Institut f\"ur 
Extraterrestrische Physik (MPE), 
National Astronomical Observatories of 
China, New Mexico State University, 
New York University, University of 
Notre Dame, Observat\'ario 
Nacional / MCTI, The Ohio State 
University, Pennsylvania State 
University, Shanghai 
Astronomical Observatory, United 
Kingdom Participation Group, 
Universidad Nacional Aut\'onoma 
de M\'exico, University of Arizona, 
University of Colorado Boulder, 
University of Oxford, University of 
Portsmouth, University of Utah, 
University of Virginia, University 
of Washington, University of 
Wisconsin, Vanderbilt University, 
and Yale University.

This research has made use of the NASA/IPAC Infrared Science Archive, which is funded by the National Aeronautics and Space Administration and operated by the California Institute of Technology.

%%%%%%%%%%%%%%%%%%%%%%%%%%%%%%%%%%%%%%%%%%%%%%%%%%
\section*{Data Availability}

The \textit{Swift}/MaNGA (SwiM) Value Added Catalogue (VAC) is publicly available at the following link: \href{https://www.sdss4.org/dr17/data\_access/value-added-catalogs/?vac\_id=swift-manga-value-added-catalog}{\texttt{https://www.sdss4.org/dr17/data\_access/value-added-\\catalogs/?vac\_id=swift-manga-value-added-catalog}}

WISE observations of the galaxies mentioned here can be publicly accessed through the NASA/IPAC Infrared Science Archive at the following link: \href{https://irsa.ipac.caltech.edu/frontpage/}{\texttt{https://irsa.ipac.caltech.edu/frontpage/}}



\bibliographystyle{mnras}
\bibliography{irxb} % if your bibtex file is called example.bib

%\begin{thebibliography}{99}
%\end{thebibliography}

%%%%%%%%%%%%%%%%%%%%%%%%%%%%%%%%%%%%%%%%%%%%%%%%%%

%%%%%%%%%%%%%%%%% APPENDICES %%%%%%%%%%%%%%%%%%%%%

%%%%%%%%%%%%%%%%%%%%%%%%%%%%%%%%%%%%%%%%%%%%%%%%%%

% Don't change these lines
\bsp	% typesetting comment
\label{lastpage}
\end{document}